\documentclass[%
preprint,
nofootinbib,
amsmath,amssymb,
aps,
prd,
]{revtex4-1}
\usepackage{setspace}
\usepackage{graphicx}
\usepackage{dcolumn}
\usepackage{bm}
\usepackage{mathtools}

\usepackage{color}
\definecolor{purple}{rgb}{0.5,0,0.5}

\usepackage[unicode=true,bookmarks=true,
bookmarksnumbered=false,bookmarksopen=false,
breaklinks=false,
pdfborder={0 0 1},
backref=false,colorlinks=true]{hyperref}
\hypersetup{pdftitle={Cosmic Equilibration: A Holographic No-Hair Theorem from the Generalized Second Law}, pdfauthor={Sean M. Carroll, Aidan Chatwin-Davies}, citecolor=blue,linkcolor=blue,urlcolor=blue,citecolor=blue}

\newtheorem{thm}{Theorem}[section]
\newtheorem{prop}[thm]{Proposition}
\newtheorem{lem}[thm]{Lemma}
\newtheorem{cor}[thm]{Corollary}

\newtheorem{conj}[thm]{Conjecture}

\newtheorem{example}[thm]{Example}

\newcommand{\pf}{\noindent {\bf Proof: }}
\newcommand{\eop}{{\hspace*{\fill}$\square$}}

\newcommand{\eoplem}{$\blacksquare$}

\newcommand{\mrm}[1]{\mathrm{#1}}

\newcommand{\Eq}[1]{Eq.~(\ref{#1})}
\newcommand{\Eqs}[2]{Eqs.~(\ref{#1}) and (\ref{#2})}
\newcommand{\Sec}[1]{Sec.~\ref{#1}}

\newcommand{\Fig}[1]{Fig.~\ref{#1}}
\newcommand{\App}[1]{App.~\ref{#1}}

\begin{document}
\baselineskip=15pt

\preprint{CALT-TH-2017-013}

\title{Cosmic Equilibration:
A Holographic No-Hair Theorem \\
from the Generalized Second Law}

\author{Sean M. Carroll and Aidan Chatwin-Davies}
\thanks{ \vspace*{-3mm}
\href{mailto:seancarroll@gmail.com}{\tt seancarroll@gmail.com},
\vspace*{-3mm}
\href{mailto:achatwin@caltech.edu}{\tt achatwin@caltech.edu}
}

\affiliation{
Walter Burke Institute for Theoretical Physics,\\
California Institute of Technology,
Pasadena, CA 91125\\
}

\begin{abstract}
In a wide class of cosmological models, a positive cosmological constant drives cosmological evolution toward an asymptotically de Sitter phase.
Here we connect this behavior to the increase of entropy over time, based on the idea that de~Sitter spacetime is a maximum-entropy state.
We prove a cosmic no-hair theorem for Robertson-Walker and Bianchi I spacetimes that admit a Q-screen (``quantum'' holographic screen) with certain entropic properties:
If generalized entropy, in the sense of the cosmological version of the Generalized Second Law conjectured by Bousso and Engelhardt, increases up to a finite maximum value along the screen, then the spacetime is asymptotically de Sitter in the future.
Moreover, the limiting value of generalized entropy coincides with the de Sitter horizon entropy.
We do not use the Einstein field equations in our proof, nor do we assume the existence of a positive cosmological constant.
As such, asymptotic relaxation to a de Sitter phase can, in a precise sense, be thought of as cosmological equilibration.

\end{abstract}

\maketitle

\tableofcontents
\newpage

\section{Introduction}

Like black holes, universes have no hair, at least if they have a positive cosmological constant $\Lambda$ \cite{Wald1983, Starobinskii1983, Barrow1987, Barrow1989, Kitada1992, Kitada1993, Bruni1995, Bruni2002, Boucher2011, Maleknejad2012}.
A cosmic no-hair theorem states that, if a cosmological spacetime obeys Einstein's equation with $\Lambda > 0$, then the spacetime asymptotically tends to an empty de~Sitter state in the future.\footnote{For a different definition of cosmic hair which more closely parallels black hole hair, see \cite{Kastor2016}.}
A more precise statement is due to Wald, who proved the following theorem \cite{Wald1983}:
\begin{thm}[Wald]
All Bianchi spacetimes (except for certain type IX spacetimes) that are initially expanding, that have a positive cosmological constant $\Lambda > 0$, and whose matter content besides $\Lambda$ obeys the strong and dominant energy conditions, tend to a de Sitter state in the future.
\end{thm}
Bianchi spacetimes are cosmologies that are homogeneous but in general anisotropic  \cite{Wald1984,Berger2014}.
For example, the metric of the 1+3-dimensional Bianchi I spacetime in comoving Cartesian coordinates is given by
\begin{equation}
ds^2 = -dt^2 + a_1^2(t) \, dx^2 + a_2^2(t) \, dy^2 + a_3^2(t) \, dz^2 .
\end{equation}
It is essentially a Robertson-Walker (RW) spacetime in which the scale factor can be different in different directions in space.
In this case, when the necessary conditions are satisfied, Wald's theorem implies that each $a_i(t)$ tends to the same de Sitter scale factor, $\exp(\sqrt{\Lambda/3} \, t)$ for a cosmological constant $\Lambda > 0$, as $t$ tends to infinity.

The intuition behind why one would expect a cosmic no-hair theorem to hold is that as space expands, the energy density of ordinary matter decreases while the density of vacuum energy remains constant.
As such, the cosmological constant eventually dominates regardless of the initial matter content and geometry, and a universe in which a positive cosmological constant is the only source of stress-energy is de~Sitter.
For Bianchi I spacetimes, one can make this intuition explicit by writing down a Friedmann equation for the average scale factor, $\bar a(t) \equiv [a_1(t) a_2(t) a_3(t)]^{1/3}$, which gives \cite[Ch.~8.6]{Kolb1994} 
\begin{equation}
\left(\frac{\dot{\bar a}(t)}{\bar a(t)}\right)^2 \propto \left( \rho_\Lambda + \rho_\mrm{matter} + \rho_\mrm{an} \right) .
\end{equation}
On the right-hand side, $\rho_\Lambda$ and $\rho_\mrm{matter}$ denote the energy densities due to the cosmological constant and matter respectively, while $\rho_\mrm{an}$ is an effective energy density due to anisotropy, similar to how one can think of spatial curvature as an effective source of stress-energy.
Crucially, $\rho_\mrm{an}$ scales at most like $\bar a^{-2}$, and so as the universe expands, only the constant contribution due to $\rho_\Lambda$ persists.
The exception to Wald's theorem is the case of a Bianchi IX spacetime (which has positive spatial curvature) whose initial matter energy density is so high that the spacetime recollapses before the cosmological constant can dominate \cite{Wald1983}.
Intuitively, we expect not only anisotropies, but also perturbative inhomogeneities to decay away at late times, though this is harder to prove rigorously \cite{Starobinskii1983,Schmidt1988,Brauer1994,Boucher2011}.
For arbitrary inhomogeneous and anisotropic cosmologies, one can always find regions that expand at least as fast as de Sitter, thus realizing a type of local no-hair theorem \cite{Kleban2016}. 
Beyond classical general relativity, various generalizations of Wald's theorem attempt to demonstrate analogous no-hair theorems for the quantum states of fields on a curved spacetime background \cite{Hollands2010,Marolf2010,Marolf2011}.

As the universe expands and the cosmological constant increases in prominence with respect to other energy sources, something else is also going on: entropy is increasing.
According to the Second Law of Thermodynamics, the entropy of any closed system (such as the universe) will increase or stay constant, at least until it reaches a maximum value.
It is interesting to ask whether there is a connection between these two results, the cosmic no-hair theorem and the Second Law.
Can the expansion of the universe toward a quiescent de~Sitter phase be interpreted as thermodynamic equilibration to a maximum-entropy state?
It is well established that de~Sitter has many of the properties of an equilibrium maximum-entropy state, including a locally thermal density matrix with a constant temperature \cite{Gibbons1977,Spradlin2001}, and the relationship between entropy and de~Sitter space has been examined from a variety of perspectives \cite{Banks:2000fe,Witten:2001kn,Dyson2002,Parikh:2004wh,Albrecht2004,Carroll:2004pn,Albrecht2009,Krishna2017,Bao:2017iye}.

In this paper we try to make one aspect of these ideas rigorous, showing that a cosmic no-hair theorem can be derived even without direct reference to Einstein's equation, simply by invoking an appropriate formulation of the Second Law.
This strategy of deducing properties of spacetime from the behavior of entropy is reminiscent of the thermodynamic and entropic gravity programs \cite{jacobson95,Padmanabhan:2009vy,Verlinde,jacobson15,Carroll:2016lku}, as well as of the gravity-entanglement connection \cite{Swingle:2009,VanRaamsdonk,Lashkari,Faulkner,Swingle:2012,graventanglement2,er-eprms,Cao:2016mst}.
Though we do not attempt to derive a complete set of gravitational field equations from entropic considerations, it is interesting that a specific spacetime can be singled out purely from the requirement that entropy increases to a maximum finite value.

To derive our theorem, we require a precise formulation of the Second Law that is applicable in curved spacetime, and that includes the entropy of spacetime itself.
A step in this direction is Bekenstein's Generalized Second Law (GSL) \cite{Bekenstein1974}.
Recall that the entropy of a black hole with area $A$ is given by $S_\mrm{BH} = A/4G$.
The GSL is the conjecture that \emph{generalized entropy}, $S_\mrm{gen}$, which is defined as the sum of the entropy of all black holes in a system as well as the ordinary thermodynamic entropy, increases or remains constant over time.
Unfortunately this form of the GSL does not immediately help us in spacetimes without any black holes.
Recently, Bousso and Engelhardt proposed a cosmological version of the GSL \cite{Bousso2015c}, building on previous work on holography \cite{Bousso1999}, apparent horizons \cite{Hayward1994,Ashtekar2002,Ashtekar2003,Ashtekar2004,Booth2005,Booth2006}, and holographic screens \cite{Bousso2015a,Bousso2015b}.
They define a version of generalized entropy on a hypersurface they call a ``Q-screen.'' A Q-screen is a quantum version of a holographic screen, which in turn is a modification of an apparent horizon.
Given a Cauchy hypersurface $\Sigma$ and a codimension-2 spatial surface with no boundary $\sigma \subset \Sigma$ that divides $\Sigma$ into an interior region and an exterior region, the generalized entropy is the sum of the area entropy of $\sigma$, i.e., its area in Planck units, and the entropy of matter in the exterior region:
\begin{equation} \label{eq:Sgendef}
S_\mrm{gen}[\sigma,\Sigma] = \frac{A[\sigma]}{4G} + S_\mrm{out}[\sigma,\Sigma].
\end{equation}
Bousso and Engelhardt's version of the GSL is the statement that generalized entropy increases strictly monotonically with respect to the flow through a specific preferred foliation of a Q-screen:
\begin{equation}
\frac{dS_\mrm{gen}}{dr} > 0 \, ,
\end{equation}
where $r$ parameterizes the foliation.
Although it is unproven in general, this version of the GSL is well motivated and known to hold in specific circumstances (the discussion of which we defer to the next section). 

In this work, we use the GSL to establish a cosmic no-hair theorem purely on thermodynamic grounds.
In an exact de~Sitter geometry, the de~Sitter horizon is a holographic screen\footnote{Pure de Sitter spacetime does not, however, satisfy the generic conditions outlined in \cite{Bousso2015b}.}, and every finite horizon-sized patch is associated with a fixed entropy that is proportional to the area of the horizon in Planck units  \cite{Sanches2016}.
We therefore conjecture that evolution toward such a state is equivalent to thermodynamic equilibration of a system with a finite number of degrees of freedom, and therefore a finite maximum entropy.
Specifically, assuming the GSL, we show that if a Bianchi I spacetime admits a Q-screen along which generalized entropy monotonically increases up to a finite maximum, then the anisotropy necessarily decays and the scale factor approaches de~Sitter behavior asymptotically in the future.
At no point do we use the Einstein field equations, nor do we assume the presence of a positive cosmological constant.
The GSL and that entropy tends to a finite maximum along the Q-screen take the logical place of these two respective ingredients.

The proof essentially consists of first showing that an approach to a finite maximum entropy heavily constrains the possible asymptotic structure of a Q-screen.
Second, we show that the spacetime must necessarily be asymptotically de Sitter (and in particular, isotropic as well) in order to admit a Q-screen with the aforementioned asymptotic structure.

The structure of the rest of this paper is as follows.
We review Q-screens and the GSL in \Sec{sec:GSL}.
In \Sec{sec:CNH-RW}, we first prove a cosmic no-hair theorem for the simpler case of RW spacetimes using the GSL.
Then, in \Sec{sec:Bianchi}, we move on to the proof for Bianchi I spacetimes, first in 1+2 dimensions to illustrate our methods, and then in 1+3 dimensions, which also illustrates how to generalize to arbitrary dimensions.
We discuss aspects of the theorems and their proofs as well as some implications in \Sec{sec:disc}.

\section{The generalized second law for cosmology}
\label{sec:GSL}

We begin by briefly reviewing Bousso and Engelhardt's conjectured Generalized Second Law (GSL).
The GSL can be thought of as a quasilocal version of Bekenstein's entropy law for black holes \cite{Bekenstein1974}, but which also applies to cosmological settings.
Moreover, the GSL is a natural semiclassical extension of Bousso and Engelhardt's area theorem for holographic screens in the same way that Bekenstein's entropy law extends Hawking's area theorem to evaporating black holes.

An early cornerstone of classical black hole thermodynamics \cite{Bardeen1973,Jacobson1996} was Hawking's area theorem: in all spacetimes which satisfy the null curvature condition, the total area of all black hole event horizons can only increase, i.e., $dA/dt \geq 0$ \cite{Hawking1972}.
Of course, the area theorem fails for evaporating black holes, the technical evasion being that they do not satisfy the null curvature condition.
Bekenstein pointed out, however, that if one instead interprets the area of the black hole event horizon as horizon entropy and includes the entropy of the Hawking radiation outside the black hole, $S_\mrm{out}$, in the total entropy budget, then the generalized entropy, $S_\mrm{gen} = A/4G + S_\mrm{out}$, increases monotonically or stays constant,  $dS_\mrm{gen}/dt \geq 0$ \cite{Bekenstein1974}.

From the perspective of trying to understand the thermodynamics of spacetime, both Hawking's and Bekenstein's results suffer from two inconveniences.
First, they are fundamentally nonlocal, since identifying event horizons requires that one know the full structure of a Lorentzian spacetime.
Second, these results only apply to black holes; it would be desirable to understand thermodynamic aspects of spacetime in other geometries as well.
These considerations motivate holographic screens \cite{Bousso2015a,Bousso2015b}, a subset of which obey a classical area theorem, as well as their semiclassical extensions called Q-screens \cite{Bousso2015c}, a subset of which are conjectured to obey an entropy theorem.
Importantly, both holographic screens and Q-screens are quasilocally defined and are known to be generic features of cosmologies in addition to black hole spacetimes.

Let us first review holographic screens.
Following the convention of Bousso and Engelhardt, here and throughout we will refer to a spacelike codimension-2 hypersurface simply as a ``surface."

Let $\sigma$ be a compact connected surface.
At every point on $\sigma$, there are two distinct future-directed null directions (or equivalently, two distinct past-directed null directions) that are orthogonal to $\sigma$: inward- and outward-directed.
The surface $\sigma$ is said to be \emph{marginal} if the expansion of the null congruence corresponding to one of these directions, say $k^\mu$, is zero everywhere on $\sigma$.
Consequently, $\sigma$ is a slice of the null sheet generated by $k^\mu$ that locally has extremal area.
This last point is particularly clear if one observes that the expansion, $\theta = \nabla_\mu k^\mu$, at a point $y \in \sigma$, can be equivalently defined as the rate of change per unit area of the area of the slice, $A[\sigma]$, when a small patch of proper area $\mathcal{A}$ is deformed along the null ray generated by $k^\mu$ at $y$ with an affine parameter $\lambda$:
\begin{equation}
\theta(y) = \lim_{\mathcal{A} \rightarrow 0} \left. \frac{1}{\mathcal{A}} \frac{dA[\sigma]}{d\lambda} \right|_y
\end{equation}
This definition is illustrated in \Fig{fig:qexp} below.

\begin{figure}[h]
\centering
\includegraphics[scale=1]{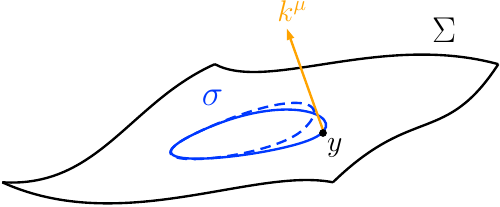}
\caption{Given a Cauchy hypersurface $\Sigma$, the surface $\sigma \subset \Sigma$ (drawn with a solid line) splits $\Sigma$ into an interior and exterior. Deformations of $\sigma$ (drawn with a dotted line) are defined by dragging $\sigma$ along the null ray generated by $k^\mu$ at any point $y \in \sigma$. More precisely, a deformation is defined by transporting a small area element $\mathcal{A} \subset \sigma$ at $y$ in the $k^\mu$ direction.}
\label{fig:qexp}
\end{figure}

A \emph{holographic screen} is a smooth codimension-1 hypersurface that can be foliated by marginal surfaces, which are then called its \emph{leaves}.
Note that while the leaves $\sigma$ are spacelike, in general a holographic screen need not have a definite character.
A marginal surface $\sigma$ is said to be \emph{marginally trapped} if the expansion of the congruence in the other null direction is negative everywhere on $\sigma$, and a \emph{future} \emph{holographic screen} is a holographic screen whose leaves are marginally trapped; marginally anti-trapped surfaces and past holographic screens are defined analogously.
Then, assuming the null curvature condition as well as a handful of mild generic conditions, Bousso and Engelhardt proved that future and past holographic screens obey the area theorem paraphrased below \cite{Bousso2015a,Bousso2015b}:

\begin{thm}[Bousso \& Engelhardt]
Let $H$ be a regular holographic screen.
The area of its leaves changes strictly monotonically under the flow through the foliation of $H$.
\end{thm}

Q-screens are related to holographic screens, but with expansion replaced by what is dubbed the ``quantum expansion.''
Let $\sigma$ again denote a compact connected surface.
The quantum expansion at a point $y \in \sigma$ in the orthogonal null direction $k^\mu$ is defined as the rate of change per unit proper area of the generalized entropy (\ref{eq:Sgendef}), i.e., the sum of both area and matter entropy, with respect to affine deformations along the null ray generated by $k^\mu$:
\begin{equation}
\Theta_k[\sigma;y] = \lim_{\mathcal{A}\rightarrow 0} \left. \frac{4G}{\mathcal{A}} \frac{d S_\mrm{gen}}{d\lambda} \right|_y
\end{equation}
Then similar to before, a \emph{quantum marginal surface} is a surface $\sigma$ such that the quantum expansion in one orthogonal null direction vanishes everywhere on $\sigma$.
Just as a marginal surface locally extremizes area along a lightsheet, a quantum marginal surface locally extremizes the generalized entropy along the lightsheet generated by $k^\mu$.

The adjective ``quantum'' can be confusing in this context. 
In this work it denotes a shift from classical general relativity, where one proves theorems about the area of surfaces, to quantum field theory on a semiclassical background, where analogous theorems refer to a generalized entropy that adds the entropy of matter degrees of freedom to such an area.
That matter entropy may be be calculated as the quantum (von Neumann) entropy of a density operator, but in the right circumstances (which we will in fact be dealing with below) it is equally appropriate to treat it as a classical thermodynamic quantity.
So here ``quantum'' should always be interpreted as ``adding an entropy term to the area of some surface,'' whether or not quantum mechanics is directly involved.

The remaining constructions have similarly parallel definitions.
A \emph{Q-screen} is a smooth codimension-1 hypersurface that can be foliated by quantum marginal surfaces.
A quantum marginal surface $\sigma$ is \emph{marginally quantum trapped} if the quantum expansion in the other null direction is negative everywhere on $\sigma$, and a \emph{future} \emph{Q-screen} is a Q-screen whose leaves are marginally quantum trapped.
Analogous definitions apply for anti-trapped surfaces and past Q-screens.
A Q-screen may be timelike, null, spacelike, or some combination thereof in different regions.
Future and past Q-screens that also obey certain generic conditions analogous to those for holographic screens are the objects that are conjectured to obey a Generalized Second Law \cite{Bousso2015c}:
\begin{conj}[Generalized Second Law]
Let $\mathcal{Q}$ be a regular future (resp. past) Q-screen. The generalized entropy of its leaves increases strictly monotonically under the past and outward (resp. future and inward) flow along $\mathcal{Q}$.
\end{conj}
Note that while the GSL remains unproven in general, it is known to hold in several examples, and it can in fact be shown to hold if one assumes the Quantum Focusing Conjecture \cite{Bousso2016}.

So far we have not said much about the precise definition of generalized entropy, so let us discuss how it is defined in more careful terms.
Our context here is quantum field theory in curved spacetime, rather than a full-blown theory of quantum gravity.
Given some spacetime, suppose that it comes equipped with a foliation by Cauchy hypersurfaces, and suppose that the spacetime's matter content is described by a density matrix $\rho(\Sigma)$ on each Cauchy hypersurface $\Sigma$.
Let $\sigma$ be a compact connected surface that divides a Cauchy hypersurface $\Sigma$ into two regions: the interior and exterior of $\sigma$.
The generalized entropy computed with respect to $\sigma$ and $\Sigma$ is then the sum of the area of $\sigma$ in Planck units and $S_\mrm{out}$, the von Neumann entropy of the reduced state of $\rho$ restricted to the exterior of $\sigma$, cf. \Eq{eq:Sgendef}.
The reduced state of $\rho$ outside $\sigma$, which we denote $\rho_\mrm{out}$, is obtained by tracing out degrees of freedom on $\Sigma$ in the interior of $\sigma$,
\begin{equation}
\rho_\mrm{out} \equiv \mrm{tr}_{\mrm{int} \, \sigma} [\rho(\Sigma)] \, ,
\end{equation}
and the Von Neumann entropy of $\rho_\mrm{out}$ is
\begin{equation}
S_\mrm{out}[\sigma,\Sigma] = - \mrm{tr}\left[ \rho_\mrm{out} \ln \rho_\mrm{out} \right] .
\end{equation}

For a general field-theoretic state, the von Neumann entropy $S_\mrm{out}[\sigma,\Sigma]$ is a formally divergent quantity.
Consequently, there is some subtlety surrounding how it should be regulated, whether through an explicit ultraviolet cutoff or via subtracting a divergent vacuum contribution \cite{Bousso2014, Bousso2015c}.
Since we will exclusively be concerned with cosmology, we will work in a regime where the matter content of the spacetime has a conserved ``thermodynamic," or coarse-grained entropy $s$ per unit comoving volume.
(Entropy per comoving volume is approximately conserved in cosmologies that do not have too much particle production \cite[Ch~3.4]{Kolb1994}.)
The von Neumann entropy of a quantum mechanical system coincides with the thermodynamic Gibbs entropy in the classical limit where the state $\rho_\mrm{out}$ has no coherence, i.e., is diagonal in the energy eigenbasis of Gibbs microstates.

We will suppose that we can take the matter contribution to the generalized entropy, which is formally given by the von Neumann entropy $S_\mrm{out}[\sigma,\Sigma]$, to be given by a coarse-grained entropy $S_{\mrm{CG}}[\sigma,\Sigma]$ in the interior of $\sigma$:
\begin{equation}
S_\mrm{out}[\sigma,\Sigma] ~ \rightarrow ~ S_\mrm{CG}[\sigma,\Sigma] = s \cdot \mrm{vol}_c[\sigma,\Sigma]
\end{equation}
Here, $\mrm{vol}_c[\sigma,\Sigma]$ denotes the comoving (coordinate) volume of $\mrm{int} \, \sigma$ on $\Sigma$.
(This approach is also taken in the examples of \cite{Bousso2015c}.)
This expression is appropriate for cosmology, where observers find themselves on the inside of Q-screens and cosmological horizons when present, as opposed to observers who remain outside of a black hole and who are unable to access the interior of the black hole's horizon.
Moreover, in the field-theoretic case where $\rho(\Sigma)$ is a pure state, then it follows that $S_\mrm{in} = S_\mrm{out}$, where $S_\mrm{in}$ is the Von Neumann entropy of $\rho_\mrm{in} \equiv \mrm{tr}_{\mrm{ext}\,\sigma}[\rho(\Sigma)]$.

The fact that each leaf of a Q-screen extremizes the generalized entropy on an orthogonal lightsheet leads to a useful method for constructing Q-screens \cite{Bousso1999}.
Given some spacetime with a foliation by Cauchy surfaces, suppose that one is also supplied with a foliation of the spacetime by null sheets with compact spatial cross-sections.
Let each null sheet be labeled by a parameter $r$, and on each null sheet, let $\sigma(r)$ be the spatial section with extremal generalized entropy, when it exists.
(Not every spacetime contains Q-screens, such as Minkowski space. But in Big Bang cosmologies, we expect both the area of, and entropy inside, a light cone to decrease in the very far past, so the generalized entropy will have an extremum somewhere.) 
It follows that each $\sigma(r)$ is a quantum marginal surface, and so if the quantum expansion has a definite sign in the other orthogonal null direction on each $\sigma(r)$, the union of these surfaces, $\mathcal{Q} = \bigcup_r \sigma(r)$, is by construction a Q-screen.

One way to generate a null foliation of a spacetime is to consider the past light cones of some timelike trajectory.
Q-screens constructed from this type of foliation will be particularly useful for our purposes.
This construction is illustrated through a worked example in Appendix~\ref{app:example}.

\section{A cosmic no-hair theorem for RW spacetimes}
\label{sec:CNH-RW}

We can used the notions reviewed above to show that spacetimes that expand and approach a constant maximum entropy along Q-screens will asymptote to de~Sitter space.
The basic idea of our proof is made clear by the simple example of a metric that is already homogeneous and isotropic, so that all we are showing is that the scale factor approaches $e^{Ht}$ for some fixed constant $H$.
The anisotropic case, considered in the next section, is considerably more complex, but the ideas are the same.

Let $\mathcal{M}$ be a Robertson-Walker (homogeneous and isotropic) spacetime with the line element
\begin{equation} \label{eq:RW}
ds^2 = -dt^2 + a^2(t) \left(d\chi^2 + \chi^2 d\Omega_{d-1}^2 \right),
\end{equation}
where $t \in (t_i,\infty)$.
Our aim is to show that if $\mathcal{M}$ admits a past Q-screen along which the generalized entropy monotonically increases up to a finite maximum value, then this alone, together with a handful of generic conditions on $\mathcal{M}$, implies that $\mathcal{M}$ is asymptotically de Sitter, or in other words, that
\begin{equation}
\lim_{t\rightarrow \infty} a(t) = e^{Ht}
\end{equation}
for some constant $H$.
In particular, we will neither make use of the Einstein field equations nor assume that there is a positive cosmological constant.

\begin{figure}[h]
\includegraphics[scale=1]{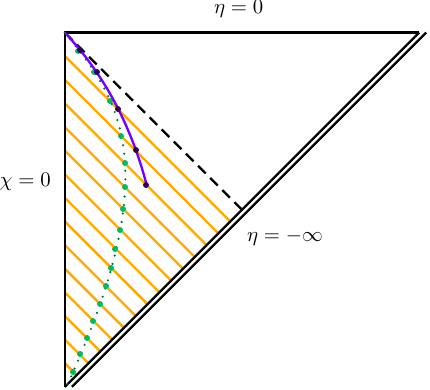}
\caption{Holographic screen and Q-screen illustrated on the Penrose diagram for a homogeneous and isotropic spacetime with a positive cosmological constant.
Null sheets that make up the foliation by past-directed light cones are shown in yellow, and the cosmological horizon is the dashed black line.
The dotted green line and large green dots are the holographic screen and its leaves respectively.
The solid purple line and large purple dots are the Q-screen and its leaves.}
\label{fig:QscreenEx}
\end{figure}

Begin by foliating $\mathcal{M}$ with past-directed light cones whose tips lie at the spatial origin $\chi = 0$, and suppose that $\mathcal{M}$ admits a past Q-screen, $\mathcal{Q}$, constructed with respect to this foliation.
In other words, suppose that each light cone has a spatial slice with extremal generalized entropy so that $\mathcal{Q}$ is the union of all of these extremal slices.
Past light cones will generically have a maximal entropy slice in cosmologies which, for example, begin with a big bang where $a(t_i) = 0$.
An example is portrayed in \Fig{fig:QscreenEx}, which shows a holographic screen and a Q-screen in a cosmological spacetime with a past null singularity and a future de~Sitter evolution; this example is explained in more detail in Appendix~\ref{app:example}.
The intuition here is that while the past-directed null geodesics that make up a light cone may initially diverge, eventually they must meet again in the past when the scale factor vanishes and space becomes singular.
Ultimately, however, we need only assume that the Q-screen exists, and we only remark on its possible origins for illustration.

Because RW spacetimes are spherically symmetric, the extremal-entropy light cone slices will be spheres, i.e., constant-$t$ slices.
If the quantum expansion vanishes in the lightlike direction along the light cone and is positive in the other lightlike direction at a single point on some test sphere, then it maintains these properties at every point on that sphere due to symmetry.
This sphere is by construction a marginally quantum anti-trapped surface, or equivalently has extremal generalized entropy on the light cone.
We therefore take the Cauchy surfaces $\Sigma$ with respect to which generalized entropy is defined to be the constant-$t$ surfaces in $\mathcal{M}$, since constant-$t$ slices of light cones are spheres.

We will also make a handful of generic assumptions about $\mathcal{M}$ and $\mathcal{Q}$ without which a cosmic no-hair theorem is not guaranteed.
Indeed, Wald's theorem does not hold in completely general cosmologies either; it requires that the spacetime is initially expanding and that its matter content satisfies the strong and dominant energy conditions.
Here, we will assume that space continues to expand for all cosmic time.\footnote{In principle, the expansion need not be monotonic, but we will find that monotonicity is implied when $\mathcal{M}$ admits a Q-screen such as $\mathcal{Q}$.}
We want to avoid cosmologies that crunch or that otherwise clearly do not admit a no-hair theorem.
We will also suppose that $\mathcal{Q}$ satisfies the generic conditions outlined in \cite{Bousso2015c}.

With these considerations in mind, the theorem that we wish to prove is the following:

\begin{thm} \label{thm:CNH-RW}
Let $\mathcal{M}$ be a RW spacetime with the line element \eqref{eq:RW} and whose matter content has constant thermodynamic entropy $s$ per comoving volume.
Suppose that $\mathcal{M}$ admits a past Q-screen, $\mathcal{Q}$, constructed with respect to a foliation of $\mathcal{M}$ with past-directed light cones that are centered on the origin, $\chi = 0$, and suppose that the Generalized Second Law holds on $\mathcal{Q}$.
Suppose that $\mathcal{M}$ and $\mathcal{Q}$ together satisfy the following assumptions:
\begin{itemize}
\item[$(a)$] $a(t) \rightarrow \infty$ as $t \rightarrow \infty$,

\item[$(b)$] $S_\mrm{gen} \rightarrow S_\mrm{max} < \infty$ along $\mathcal{Q}$.
\end{itemize}
Then, $\mathcal{M}$ is asymptotically de Sitter and the scale factor $a(t)$ approaches $e^{Ht}$, where $H$ is a constant.
\end{thm}

\pf For convenience we work in $d=3$ spatial dimensions, but the generalization to arbitrary dimensions is straightforward. As discussed above, the leaves of $\mathcal{Q}$ are spheres.
Letting the leaves be labeled by some parameter $r$, the generalized entropy is then given by
\begin{equation}
\label{eq:sgenRW}
S_\mrm{gen}[\sigma(r),\Sigma(r)] \equiv S_\mrm{gen}(r) = \frac{\pi}{G} \chi^2(r) a^2(t(r)) + \tfrac{4}{3} \pi \chi^3(r) s \, .
\end{equation}
The hypersurface $\Sigma(r)$ is the constant-$t(r)$ surface in which the leaf $\sigma(r)$ is embedded, and $\chi(r)$ denotes the radius of the leaf.

First, we need to establish that $\mathcal{Q}$ extends out to future timelike infinity.
In principle, $\mathcal{Q}$ could become spacelike and consequently not extend beyond some time $t$ (or in other words, $t(r)$ could have some finite maximum value), but it turns out that this does not happen.

Recall the property of Q-screens that generalized entropy is extremized on each leaf with respect to lightlike deformations.
Here we may write
\begin{equation}
k^\mu \partial_\mu S_\mrm{gen} = 0 \, ,
\label{eq:sgendiv}
\end{equation}
where $k^\mu = (a(t), -1, 0, 0)$ is the lightlike vector that is tangent to the light cone and with respect to which $S_\mrm{gen}$ is extremal.
(Any point $x^\mu$ belongs to a unique sphere on a past-directed light cone and may therefore be associated with a particular value of $S_\mrm{gen}$.
This lets us define the partial derivative in \Eq{eq:sgendiv} above.)
The deformation corresponds to dragging the leaf $\sigma(r)$ up and down the light cone, and by construction $S_\mrm{gen}(r)$ is extremal on the leaf $\sigma(r)$.
Note that in more general settings we should consider deformations with respect to null \emph{geodesics}, since the null generators of the light cone could have different normalizations at different points on $\sigma(r)$.
Or, in other words, the geometry of the leaf $\sigma(r)$ could change as it is dragged by some fixed affine amount along the light cone.
Here, however, the spherical symmetry of RW ensures that the null generators on $\sigma(r)$ all have the same normalization, so that $k^\mu$ as defined above is proportional to the null generators everywhere on $\sigma(r)$.

Writing out the partial derivatives, (\ref{eq:sgendiv}) becomes
\begin{align}
0 &= (a \, \partial_t - \partial_\chi) \left( \frac{\pi}{G} \chi^2 a^2 + \tfrac{4}{3} \pi \chi^3 s \right)\nonumber \\
&= \frac{2\pi}{G} \chi^2 a^2 \dot a - \frac{2\pi}{G} \chi a^2 - 4 \pi \chi^2 s .
\end{align}
(One must be careful to distinguish between the coordinate $t$ and the value $t(r)$ which labels leaves in the Q-screen.)
If $\chi \neq 0$, then it follows that
\begin{equation} \label{eq:Qexist}
\frac{1}{\chi} = \dot a(t) - \frac{2Gs}{a^2(t)} \, .
\end{equation}
\Eq{eq:Qexist} lays out the criterion for when there is a leaf in a constant-$t$ slice; when the right side is finite and positive, then there must be a leaf in that slice.

Observe that the right side of \Eq{eq:Qexist} does not diverge for any finite $t > t_i$ since $a(t)$ is defined for all $t \in [t_i,\infty)$ and only diverges in the infinite $t$ limit by assumption.
Furthermore, if the right side is nonzero and positive for some time $t_\mrm{time}$ (and consequently there is a leaf $\sigma(r)$ in the $t(r) = t_\mrm{time}$ slice), then the right side cannot approach zero, since this would cause the radius of subsequent leaves to grow infinitely large, which contradicts the assumption that $S_\mrm{gen}$ remains finite.
Therefore, if $\mathcal{Q}$ has a leaf at \emph{some} time, then \Eq{eq:Qexist} shows that $\mathcal{Q}$ must have leaves in \emph{all} future slices.
$\mathcal{Q}$ is therefore timelike and extends out to future timelike infinity.\footnote{Alternatively, we could instead replace Assumption~$(a)$ with the assumption that $\mathcal{Q}$ is timelike and extending out to future timelike infinity and argue that $a \rightarrow \infty$. The arguments given here show that these two points are logically equivalent.}
Furthermore, that the right side of \Eq{eq:Qexist} cannot vanish immediately implies that $\dot a > 2Gs/a^2 > 0$ for $t > t_\mrm{time}$, so that the expansion must be monotonic.

Because $\mathcal{Q}$ is timelike, we can label each leaf by the constant-$t_1$ surface in which it lies, i.e., let the parameter $r$ be a time $t_1$ (subscripted as such to distinguish it from the coordinate $t$).
Referring to \Eq{eq:sgenRW}, since $a(t)$ grows without bound by assumption, it must be that $\chi(t_1)$ decreases at least as fast as $a^{-1}(t_1)$ in order for the area term in $S_\mrm{gen}$ to remain finite (as it must, since by hypothesis $S_\mrm{gen} \leq S_\mrm{max}$).
The matter entropy term therefore becomes irrelevant in the asymptotic future, and so that $S_\mrm{gen} \rightarrow S_\mrm{max}$, it must be that
\begin{equation} \label{eq:chiRW}
\chi(t_1) \rightarrow \sqrt{\frac{G S_\mrm{max}}{\pi}} \frac{1}{a(t_1)}
\end{equation}
as $t_1 \rightarrow \infty$.

Next, rearrange \Eq{eq:Qexist} to solve for $\dot a$.
Using the asymptotic form for $\chi(t_1)$ in \Eq{eq:chiRW}, to leading order in $a$ we find that
\begin{equation} \label{eq:RWasymp}
\dot a \rightarrow \sqrt{\frac{\pi}{G S_\mrm{max}}} a \;\; + \;\; \mrm{(subleading)} \, .
\end{equation}
Therefore, it follows that $a(t) \rightarrow e^{Ht}$ as $t \rightarrow \infty$, where $H = (\pi/GS_\mrm{max})^{1/2}$, demonstrating that the metric approaches the de~Sitter form, as desired.
The entropy $S_\mrm{max} = \pi/GH^2$ coincides with the usual de Sitter horizon entropy.
\eop

~

We close this section by briefly remarking that the result above extends straightforwardly to open and closed RW spacetimes as well.

\begin{cor}
More generally, the result of Theorem~\ref{thm:CNH-RW} applies to a RW spacetime $\mathcal{M}$ of any spatial curvature, i.e., with the line element
\begin{equation}
ds^2 = -dt^2 + a^2(t) \left(d\chi^2 + f^2(\chi) d\Omega_{d-1}^2 \right)
\end{equation}
where
\begin{equation}
f(\chi) = \left\{
\begin{array}{lll}
\sin \chi & ~ \chi \in [0,\pi] & ~ \mrm{(closed)} \\
\chi & ~ \chi \in [0,\infty) & ~ \mrm{(flat)} \\
\sinh \chi & ~ \chi \in [0,\infty) & ~ \mrm{(open)}
\end{array}
\right. \, .
\end{equation}
\end{cor}

\pf
The overall proof technique is the same as in the proof of Theorem~\ref{thm:CNH-RW}.
Working in 1+3 dimensions, in the more general case, the generalized entropy of the leaves of $\mathcal{Q}$ is given by
\begin{equation}
S_\mrm{gen}[\sigma(r),\Sigma(r)] \equiv S_\mrm{gen}(r) = \frac{\pi}{G} f^2(\chi(r)) a^2(t(r)) + v(\chi(r)) s \, .
\end{equation}
When $\mathcal{M}$ is closed, the comoving volume $v(\chi)$ is given by $v(\chi) = 2\pi (\chi - \sin \chi \cos \chi)$, and when $\mathcal{M}$ is open, $v(\chi)$ is given by $v(\chi) = 2\pi (\sinh \chi \cosh \chi - \chi)$.
Consequently, the condition $k^\mu \partial_\mu S_\mrm{gen} = 0$, which determines when there is a leaf in the constant-$t$ hypersurface, gives
\begin{equation} \label{eq:Qexistgen}
\frac{1}{f^2(\chi)} = \left( \dot a(t) - \frac{2Gs}{a^2(t)}  \right)^2 + k \, ,
\end{equation}
where $k = +1, 0,$~or~$-1$ if $\mathcal{M}$ is respectively closed, flat, or open.
Here as well, if there is a leaf at some time $t_\mrm{time}$ so that the right-hand side of \Eq{eq:Qexistgen} is nonzero, then there are leaves in all subsequent constant-$t$ slices, since the finiteness of $S_\mrm{gen}$ demands that the right-hand side cannot approach zero.
Therefore, $\mathcal{Q}$ extends out to future timelike infinity.

For the general case, the condition in \Eq{eq:chiRW} that $S_\mrm{gen} \rightarrow S_\mrm{max}$ reads\footnote{A minor technical point worth noting is that the condition in \Eq{eq:chiRWgen} is not identically equivalent to the condition $\chi(t_1) \rightarrow \sqrt{GS_\mrm{max}/\pi}/a(t_1)$ when $\mathcal{M}$ is closed. In this case, $\chi(t_1) \rightarrow \pi - \sqrt{GS_\mrm{max}/\pi}/a(t_1)$ is also admissible.}
\begin{equation} \label{eq:chiRWgen}
f(\chi(t_1)) \rightarrow \sqrt{\frac{GS_\mrm{max}}{\pi}} \frac{1}{a(t_1)} \, .
\end{equation}
Upon substituting \Eq{eq:chiRWgen} into \Eq{eq:Qexistgen} (and taking the positive root, since $\mathcal{M}$ is expanding), we recover \Eq{eq:RWasymp}, and so the rest of the proof follows as before.
\eop

\section{A cosmic no-hair theorem for Bianchi I spacetimes}
\label{sec:Bianchi}

In a RW spacetime, we demonstrated that the existence of a Q-screen along which entropy monotonically increases to a finite maximum implies that the scale factor tends to the de Sitter scale factor far in the future.
Now we will go one step further and show that in the case where the cosmology is allowed to be anisotropic, similar assumptions imply that any initial anisotropies decay at late times as well.
Specifically, we will prove a cosmic no-hair theorem for Bianchi I spacetimes.
The calculations in the proof for Bianchi I spacetimes are more involved than the RW case, so we will begin with a proof in 1+2 dimensions, where the anisotropy only has one functional degree of freedom.
We will then generalize to 1+3 dimensions, which also makes apparent how to generalize to arbitrary dimensions.

\subsection{1+2 dimensions}

Let $\mathcal{M}$ be a Bianchi I spacetime in 1+2 dimensions with the line element
\begin{equation} \label{eq:bianchiI3d}
ds^2 = -dt^2 + a_1^2(t) \, dx^2 + a_2^2(t) \, dy^2
\end{equation}
where $t \in (t_i, \infty)$.
Once again foliate $\mathcal{M}$ with past-directed light cones whose tips lie at $x = y = 0$, and suppose that $\mathcal{M}$ admits a past Q-screen $\mathcal{Q}$, constructed with respect to this foliation, together with an accompanying foliation by Cauchy hypersurfaces.
Our aim is to show that if generalized entropy tends to a finite maximum along $\mathcal{Q}$, then the GSL implies that $a_1(t), a_2(t) \rightarrow e^{Ht}$ as $t \rightarrow \infty$ for some constant $H$.

Here as well we will assume that space expands for all time, with $a_1(t), a_2(t) \rightarrow \infty$ as $t \rightarrow \infty$.
We will also further assume that $\mathcal{Q}$ is timelike and extends out to future timelike infinity past some time $t_\mrm{time}$.
We suspect that it might be possible to show that this latter property follows from the assumption that $a_1(t)$ and $a_2(t)$ grow without bound, as in the case of a RW spacetime, but we do not know of a straightforward way to show this.

We will assume that generalized entropy is \emph{globally} maximized on each light cone by the corresponding screen leaf (as opposed to only assuming local extremality).
In other words, we will assume that there are no other slices of each light cone whose generalized entropy is larger than that of the screen leaf.
This property of leaves is certainly true when the Quantum Focusing Conjecture (QFC) holds \cite{Bousso2016}.
Moreover, the GSL is provably true when the QFC holds.

The QFC is the conjecture that the quantum expansion of a null congruence is nonincreasing along the congruence.
In symbols, for a null congruence generated by $k^\mu$ with an affine parameter $\lambda$ on a given null ray, the QFC reads
\begin{equation} \label{eq:QFC}
\frac{d \Theta_k}{d\lambda} \leq 0 \, .
\end{equation}
The QFC is the semiclassical analogue of classical focusing, $d\theta/d\lambda \leq 0$, which holds when the null curvature condition holds.
In particular, \Eq{eq:QFC} makes it clear that if a light cone slice locally maximizes generalized entropy with respect to deformations on the light cone, then it is also the unique global maximum.
A leaf $\sigma$ that locally maximizes generalized entropy obeys $\Theta_k[\sigma,y] = 0$ for all $y \in \sigma$.
Therefore, if $\Theta_k$ is nonincreasing on the light cone\footnote{The pathological case of $d\Theta_k/d\lambda = 0$ on a subset of the congruence with nonzero measure is ruled out by appropriate genericity conditions.}, there are no deformations of $\sigma$ that lead to a larger generalized entropy, and so $\sigma$ defines a globally maximal generalized entropy.
It is interesting to explore ways in which this assumption about global maximality of generalized entropy can be relaxed, which we shall do after the proof of the no-hair theorem.

\begin{figure}[h]
\centering
\includegraphics[scale=1]{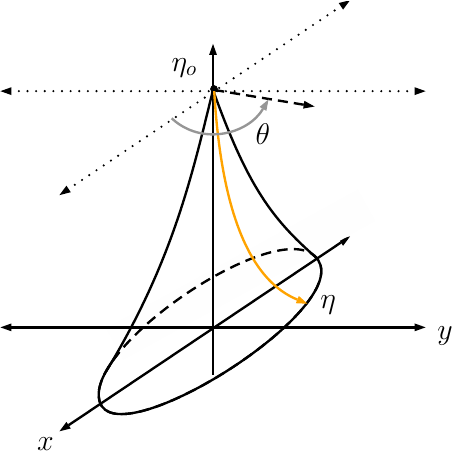}
\caption{Conformal light cone coordinates. At the conformal time $\eta_o$, fire a past-directed null geodesic (shown in yellow) from the origin at an initial angle $\theta$ relative to the positive $x$-axis and follow it until the conformal time $\eta$.}
\label{fig:clc}
\end{figure}

Next, we introduce \emph{conformal light cone coordinates} \cite{Saunders1969,Fleury2016}, which are more convenient coordinates  to work in when dealing with anisotropy.
First, observe that we may rewrite the line element \eqref{eq:bianchiI3d} as
\begin{equation} \label{eq:bianchiI3d_ab}
ds^2 = -dt^2 + a^2(t)\left[e^{2b(t)} dx^2 + e^{-2b(t)} dy^2 \right]
\end{equation}
with $a_1(t) = a(t) \, e^{b(t)}$ and $a_2(t) = a(t) \, e^{-b(t)}$ \cite{Berger2014}.
In this parameterization, the ``volumetric scale factor'' $a(t)$ describes the overall expansion of space while $b(t)$ characterizes the anisotropy.
Next, make the coordinate transformation to conformal time defined by $dt = \pm a(\eta)~d\eta$ so that the line element \eqref{eq:bianchiI3d_ab} reads
\begin{equation} \label{eq:conformallineelement}
ds^2 = a^2(\eta) \left[ -d\eta^2 + e^{2b(\eta)} dx^2 + e^{-2b(\eta)} dy^2 \right].
\end{equation}
Choose the sign of $\eta$ so that $\eta(t)$ is a monotonically increasing function of $t$, and denote the limiting value of $\eta(t)$ as $t \rightarrow \infty$ by $\eta_\infty$.
Conformal light cone coordinates are then defined by the following coordinate transformation:
\begin{align}
\label{eq:comovingLC3D_1} x(\eta,\eta_o,\theta) &= \cos \theta \int_{\eta}^{\eta_o}   \frac{e^{-2b(\zeta)}}{ \sqrt{ \cos^2 \theta ~ e^{-2b(\zeta)} + \sin^2 \theta~e^{2b(\zeta)} } }~d\zeta \\
\label{eq:comovingLC3D_2} y(\eta,\eta_o,\theta) &= \sin \theta \int_{\eta}^{\eta_o}  \frac{e^{2b(\zeta)}}{ \sqrt{ \cos^2 \theta ~ e^{-2b(\zeta)} + \sin^2 \theta~e^{2b(\zeta)} } }~d\zeta
\end{align}
The point with coordinates $(\eta,\eta_o,\theta)$ is reached by firing a past-directed null geodesic from the spatial origin $x = y = 0$ at an angle $\theta \in [0,2\pi)$ counterclockwise relative to the $x$-axis at conformal time $\eta_o$ and following the light ray in the past down to the conformal time $\eta$ (\Fig{fig:clc}). 
Note that while $\eta$ is a timelike coordinate, $\eta_o$ acts as a radial coordinate at each $\eta$.

\begin{figure}[h]
\centering
\includegraphics[scale=1]{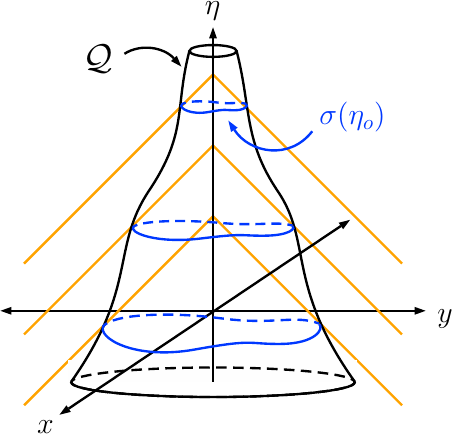}
\caption{A Q-screen $\mathcal{Q}$ (the solid black hypersurface) constructed with respected to a foliation by past-directed light cones (sketched in yellow). Each leaf $\sigma(\eta_o)$ (shown in blue) is labelled by the value of $\eta_o$ where the tip of its parent light cone sits.
}
\label{fig:Qscreen}
\end{figure}

The surfaces of constant $\eta_o$ are precisely the past-directed light cones with respect to which $\mathcal{Q}$ is constructed.
We can therefore label the leaves $\sigma$ of $\mathcal{Q}$ by the values of $\eta_o$ corresponding to the light cones on which they lie (\Fig{fig:Qscreen}):
\begin{equation}
\mathcal{Q} = \bigcup_{\eta_o}~\sigma(\eta_o)
\end{equation}
Similarly, label the Cauchy hypersurfaces with respect to which each leaf is defined by $\Sigma(\eta_o)$.
In various instances, it will be useful to use another coordinate,
\begin{equation}
  \chi = \eta_o - \eta ,
\end{equation}
which may be thought of as a comoving radius in a sense that will be made precise later.
We will also sometimes work in the coordinates $(\eta,\chi,\theta)$ or $(\chi,\eta_o,\theta)$ in addition to the conformal light cone coordinates $(\eta,\eta_o,\theta)$.

The no-hair theorem that we will prove is as follows:
\begin{thm} \label{thm:CNH1+2}
Let $\mathcal{M}$ be a Bianchi I spacetime with the line element \eqref{eq:bianchiI3d} and whose matter content has constant thermodynamic entropy $s$ per comoving volume.
Suppose that $\mathcal{M}$ admits a past Q-screen $\mathcal{Q}$, with globally maximal entropy leaves constructed with respect to a foliation of $\mathcal{M}$ with past-directed light cones that are centered on the origin, $x = y = 0$.
Suppose that the Generalized Second Law holds on $\mathcal{Q}$ and that $\mathcal{M}$ and $\mathcal{Q}$ together satisfy the following assumptions:
\begin{itemize}
\item[$(i)$] $a_1(t), a_2(t) \rightarrow \infty$ as $t \rightarrow \infty$,
\item[$(ii)$] $\mathcal{Q}$ is timelike past some $t_\mrm{time}$ and extends out to future timelike infinity
\item[$(iii)$] $\dot a_1(t), \dot a_2(t) > 0$ after some $t_\mrm{mono}$,
\item[$(iv)$] $S_\mrm{gen} \rightarrow S_\mrm{max} < \infty$ along $\mathcal{Q}$.
\end{itemize}
Then, $\mathcal{M}$ is asymptotically de Sitter and the scale factors $a_1(t)$ and $a_2(t)$ approach $C_1 e^{Ht}$ and $C_2 e^{Ht}$, respectively, where $H$, $C_1$, and $C_2$ are constants.
\end{thm}
{\bf Notes:} To obtain a manifestly isotropic metric, rescale the coordinates $x$ and $y$ by $C_1$ and $C_2$, i.e., set $X = C_1 x$ and $Y = C_2 y$.
Then, the line element \eqref{eq:bianchiI3d} asymptotically reads $ds^2 \rightarrow -dt^2 + e^{2Ht}(dX^2 + dY^2)$.
Also note that we have introduced an additional assumption compared to the RW case: Assumption~$(iii)$, that $a_1(t)$ and $a_2(t)$ grow monotonically past some time $t_\mrm{mono}$.
Finally, also note that in terms of $a(\eta)$ and $b(\eta)$, Assumption $(i)$ becomes:
\begin{itemize}
\item[$(i^\prime)$] $a(\eta) \rightarrow \infty$ as $\eta \rightarrow \eta_\infty$ and $a(\eta) e^{\pm b(\eta)} \rightarrow \infty$.
\end{itemize}
In terms of $a(\eta)$ and $b(\eta)$, the theorem is established by showing that $a(\eta) \rightarrow -1/H\eta$ and $b(\eta) \rightarrow B$ as $\eta \rightarrow 0^-$ (and also that $\eta_\infty = 0$) for some constant $B$.

~

\pf
The proof can be broken down into three parts.
First, we show that, asymptotically, $\mathcal{Q}$ squeezes into the comoving coordinate origin.
Second, we use this asymptotic squeezing behaviour to demonstrate that the volumetric scale factor $a(\eta)$ tends to the de Sitter scale factor.
Finally, we show that the asymptotic behaviour of $a(\eta)$ and Assumption $(iii)$ together imply that anisotropy decays.

\subsubsection{Showing that $\mathcal{Q}$ squeezes into the coordinate origin $\chi=0$ as $\eta \rightarrow \eta_\infty$.}

Consider the leaves of $\mathcal{Q}$ and work in $\tilde x^\mu = (\eta,\eta_o,\theta)$ coordinates.
On the light cone whose tip is at $\eta_o$, each leaf $\sigma(\eta_o)$ is a closed path parameterized by
\begin{equation}
\tilde x^\mu(u; \eta_o) = (\eta(u; \eta_o), \eta_o, u) \qquad u \in [0,2\pi).
\end{equation}
Our first task is to show that $\chi(u;\eta_o) \equiv \eta_o - \eta(u;\eta_o)$ tends to zero for all values of $u$ as $\eta \rightarrow \eta_\infty$.
We will do so through a proof by contradiction.

\begin{figure}[h]
\centering
\includegraphics[scale=1]{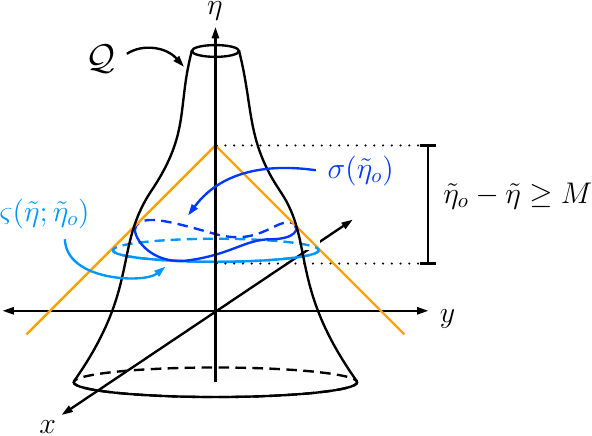}
\caption{The leaf $\sigma(\tilde \eta_o)$ and the constant-$\tilde \eta$ slice, $\varsigma(\tilde \eta; \tilde \eta_o)$, of its parent light cone.}
\label{fig:bound_slice}
\end{figure}

Suppose to the contrary that $\mathcal{Q}$ never squeezes into the comoving coordinate origin.
That is, suppose that there exists $M > 0$ such that, given any $\eta_o > \eta_\mrm{time}$, one can find values $\tilde \eta_o > \eta_o$ and $\tilde u$ such that $\chi(\tilde u; \tilde \eta_o) \geq M$.
Let $\tilde \eta \equiv \eta(\tilde u; \tilde \eta_o)$ and consider the constant $\eta = \tilde \eta$ slice of the light cone whose tip is at $\tilde \eta_o$ (\Fig{fig:bound_slice}).
Denote this (co-dimension 2) surface by $\varsigma(\tilde \eta ; \tilde \eta_o)$, and denote the (co-dimension 1) hypersurface of constant-$\tilde \eta$ by $X(\tilde \eta)$.
Since the generalized entropy of the leaf $\sigma(\tilde \eta_o)$ is globally maximal on this light cone by assumption, it must follow that
\begin{equation} \label{eq:SgenLowerBound}
S_\mrm{gen}[\sigma(\tilde \eta_o),\Sigma(\tilde \eta_o)] \geq S_\mrm{gen}[\varsigma(\tilde \eta ; \tilde \eta_o), X(\tilde \eta)] \geq \frac{A[\varsigma(\tilde \eta ; \tilde \eta_o)]}{4G} \, ,
\end{equation}
where the last inequality follows because $S_\mrm{gen}$ is always greater than or equal to just the area term.
Our basic strategy will be to show that $A[\varsigma(\tilde \eta ; \tilde \eta_o)]$ diverges as $\tilde \eta_o \rightarrow \eta_\infty$, which contradicts the assumption $(iv)$ that $S_\mrm{gen}$ must remain finite on $\mathcal{Q}$.

To do this, let us compute the proper area $A[\varsigma(\tilde \eta ; \tilde \eta_o)]$.
In three dimensions, the induced metric on a surface of constant $\eta$ and $\eta_o$ has only a single component, given by
\begin{equation}
\gamma = \frac{\partial x^\mu}{\partial \theta} \frac{\partial x^\nu}{\partial \theta} g_{\mu\nu} = a^2(\eta) \left[ e^{2b(\eta)} \left( \frac{\partial x}{\partial \theta} \right)^2 + e^{- 2b(\eta)} \left( \frac{\partial y}{\partial \theta} \right)^2 \right] \equiv a^2(\eta)~\tilde \gamma \, ,
\label{eq:inducedmetric}
\end{equation}
where the coordinate partial derivatives read\footnote{A Maple worksheet which implements the calculations in this article is available through the online repository \cite{Chatwin-Davies2018}.}
\begin{align*}
\frac{\partial x}{\partial \theta} &= \int_\eta^{\eta_o} \frac{- \sin \theta}{\left( \cos^2 \theta~e^{-2 b(s)} + \sin^2 \theta~e^{2b(s)}  \right)^{3/2}}~ds \\
\frac{\partial y}{\partial \theta} &= \int_\eta^{\eta_o} \frac{\cos \theta}{\left( \cos^2 \theta~e^{-2 b(s)} + \sin^2 \theta~e^{2b(s)}  \right)^{3/2}}~ds .
\end{align*}
It follows that the area of this surface is
\begin{equation} \label{eq:AconstExact}
A(\eta,\eta_o) = \int_0^{2\pi} \sqrt{\gamma}~d\theta = a(\eta) \int_0^{2\pi} \sqrt{\tilde \gamma}~d\theta \, .
\end{equation}
It is fairly straightforward to place a lower bound on this area:
\begin{align*}
A(\eta,\eta_o) &\geq a(\eta) e^{b(\eta)} \int_0^{2\pi} \left| \frac{\partial x}{\partial \theta} \right|~d\theta \\
&= a(\eta) e^{b(\eta)} \int_\eta^{\eta_o} ds \int_0^{2\pi} d\theta \frac{\left| \sin \theta \right|}{\left( \cos^2 \theta~e^{-2 b(s)} + \sin^2 \theta~e^{2b(s)}  \right)^{3/2}} \\
&= 4 \, a(\eta) e^{b(\eta)} \int_\eta^{\eta_o} ds~e^{-b(s)}
\end{align*}
One arrives at a similar expression using $\partial y/\partial \theta$.
Note that in the middle line above, we were able to bring the absolute value into the integrand of $\partial x/\partial \theta$ because it has a definite sign for any given $\theta$.
Then, if $e^{-b(s)}$ is minimized at $s = \eta_m \in [\eta,\eta_o]$, it follows that
\begin{equation}
A(\eta,\eta_o) \geq 4 \, a(\eta) e^{b(\eta)}e^{-b(\eta_m)} (\eta_o - \eta) \geq 4 \, a(\eta) (\eta_o - \eta) .
\end{equation}
Applied to our surface $\varsigma(\tilde \eta; \tilde \eta_o)$, for which $\tilde \eta_o - \tilde \eta \geq M$, the bound reads
\begin{equation}
A[\varsigma(\tilde \eta; \tilde \eta_o)] \equiv A(\tilde \eta,\tilde \eta_o) \geq 4 M a(\tilde \eta) \, ,
\end{equation}
which diverges as $\tilde \eta_o$ and $\tilde \eta$ are chosen arbitrarily large.
We therefore have the contradiction that we sought, and so the leaves of the Q-screen must squeeze into the comoving coordinate origin in the asymptotic future.

\subsubsection{Showing that $a(\eta)$ is asymptotically de Sitter.}
\label{sec:asympdS}

Now we turn our attention to calculating $S_\mathrm{gen}[\sigma(\eta_o), \Sigma(\eta_o)]$ itself, and using its asymptotic properties as $\eta_o \rightarrow \eta_\infty$ to demonstrate that $a(\eta) \rightarrow -1/H\eta$ for a constant $H$ with $\eta_\infty = 0$.
First, we we will argue that the matter entropy term, which we assume can be calculated using the coarse-grained entropy $S_\mrm{CG}[\sigma(\eta_o),\Sigma(\eta_o)]$, vanishes asymptotically in the future.
To this end, let us prove the following useful lemma about constant-$\eta$ slices of light cones when $\chi = \eta_o - \eta$ is infinitesimally small:

\begin{lem} \label{lem:constSlice}
Let $\varsigma(\eta; \eta+\chi)$ be the constant-$\eta$ slice of the past-directed light cone whose tip is at $\eta_o = \eta + \chi$.
The generalized entropy defined by this slice is given by
\begin{equation}
S_\mathrm{gen}[\varsigma(\eta;\eta+\chi),X(\eta)] = \frac{A(\eta,\eta+\chi)}{4G} + c_g(\eta,\chi) \chi^2 s,
\end{equation}
where $A(\eta,\eta+\chi)$ is given by
\begin{equation} \label{eq:A3Dasymptotic}
A(\eta,\eta+\chi) = a(\eta) \cdot \left[ 2\pi \chi + O(\chi^3) \right],
\end{equation}
and $c_g(\eta,\chi)$ is some $O(1)$ geometric factor due to anisotropy that does not depend on $a(\eta)$.
\end{lem}

\pf
First we justify the parameterization of the coarse-grained entropy $S_\mathrm{CG} = c_g(\eta,\chi)\chi^2 s$.
In the coordinates of the metric \eqref{eq:conformallineelement}, $S_\mathrm{CG}$ is given by
\begin{equation}
S_\mathrm{CG}[\varsigma(\eta;\eta+\chi),X(\eta)] = s \cdot \mrm{vol}_c[\varsigma(\eta;\eta+\chi),X(\eta)] = s \iint_{\mathrm{int} \, \varsigma} dx \, dy \, ,
\end{equation}
where $\mathrm{int} \, \varsigma(\eta;\eta+\chi)$ denotes the region on $X(\eta)$ inside $\varsigma(\eta;\eta+\chi)$.
In terms of the coordinates $(\eta,\chi,\theta)$, $S_\mathrm{CG}$ is
\begin{equation}
S_\mathrm{CG}[\varsigma(\eta;\eta+\chi),X(\eta)] \equiv S_\mathrm{CG}(\eta,\chi) = s \int_0^\chi \int_0^{2\pi} \left| \frac{\partial (x,y)}{\partial (\chi^\prime,\theta)} \right| ~ d\theta \, d\chi^\prime \, .
\end{equation}
Formally, the Jacobian can be calculated from the coordinate transformation \eqref{eq:comovingLC3D_1}-\eqref{eq:comovingLC3D_2} above.
Expanding in powers of $\chi$, one finds that
\begin{equation}
S_\mathrm{CG}(\eta,\chi) = s \cdot \left( \pi \chi^2 + \frac{\pi}{8} b^\prime(\eta)^2 \chi^4 \right) + O(\chi^5) \, .
\end{equation}
Therefore, we can simply define the function $c_g(\eta,\chi)$ to be the function
\begin{equation}
c_g(\eta,\chi) \equiv \frac{S_\mathrm{CG}(\eta,\chi)}{\chi^2 s} = \pi + \frac{\pi}{8} b^\prime(\eta)^2 \chi^2 + O(\chi^3) \, .
\end{equation}
The function $c_g(\eta,\chi)$ is $O(\chi^0)$ by construction, and from the coordinate transformation \eqref{eq:comovingLC3D_1}-\eqref{eq:comovingLC3D_2}, in which $a(\eta)$ never appears, we see that $c_g$ cannot depend on $a(\eta)$, as claimed.

The expansion of $A(\eta,\eta+\chi)$ for small $\chi$ follows from expanding $\sqrt{\tilde \gamma}$ in \Eq{eq:AconstExact} in powers of $\chi$ and then integrating.
Note that \Eq{eq:A3Dasymptotic} demonstrates the sense in which $\chi$ is a comoving radius (at least for small values of $\chi$). \eoplem

~

\begin{figure}
\centering
\includegraphics[scale=1]{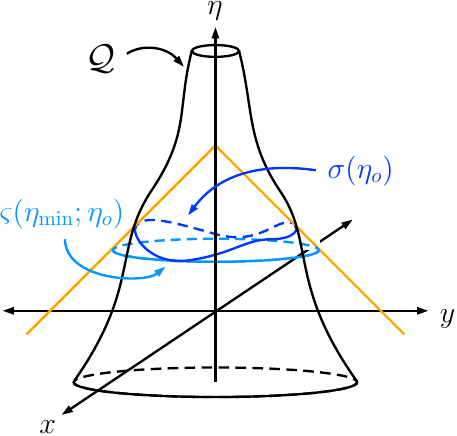}
\caption{Given a leaf $\sigma(\eta_o)$, the constant-$\eta = \eta_\mrm{min}$ slice of its parent light cone is the surface $\varsigma(\eta_\mrm{min};\eta_o)$.}
\label{fig:const_slice}
\end{figure}

We can use the result of Lemma~\ref{lem:constSlice} to show that $S_\mrm{CG}[\sigma(\eta_o),\Sigma(\eta_o)]$ vanishes asymptotically in the future.
Given a leaf $\sigma(\eta_o)$, let $\eta_\mrm{min}$ be the minimum value attained by $\eta(u;\eta_o)$:
\begin{equation}
\eta_\mrm{min} = \min_u \left\{ \eta(u; \eta_o) \right\}
\end{equation}
Consider the constant-$\eta_\mrm{min}$ slice of the light cone whose tip is at $\eta_o$, which we label by $\varsigma(\eta_\mrm{min};\eta_o)$ (\Fig{fig:const_slice}).
The comoving volume of $\sigma(\eta_o)$ is contained within the comoving volume of $\varsigma(\eta_\mrm{min};\eta_o)$, which, according to Lemma~\ref{lem:constSlice}, vanishes in the asymptotic future limit.
Therefore, the comoving volume of $\sigma(\eta_o)$ vanishes as well, so $S_\mrm{CG}[\sigma(\eta_o),\Sigma(\eta_o)]$ vanishes asymptotically in the future.

Next we investigate the asymptotic behaviour of $A[\sigma(\eta_o)]$.
For this part of the proof, we will work in the coordinates $(\chi,\eta_o,\theta)$.
In these coordinates, the leaf $\sigma(\eta_o)$ is parameterized by some path $\tilde x^\mu(u) = (\chi(u;\eta_o),\eta_o,u)$ with $\eta_o$ held constant and $0 \leq u < 2\pi$.
In the future when $S_\mrm{CG}$ becomes negligible, this path is the maximal area (also known as length in 1+2 dimensions) path on the light cone whose tip is at $\eta_o$, and so $A[\sigma(\eta_o)]$ satisfies
\begin{equation}
\frac{\delta A[\sigma(\eta_o)]}{\delta \chi(u;\eta_o)} = 0 \, .
\end{equation}
In principle, one can therefore solve the Euler-Lagrange problem above to obtain the path $\chi(u; \eta_o)$ and hence also the maximal area $A[\sigma(\eta_o)]$.

The tangent to the path is $t^\mu = d\tilde x^\mu/du = (\dot \chi(u; \eta_o), 0, 1)$ (where a dot denotes a derivative with respect to the parameter $u$).
Therefore, the area of $\sigma(\eta_o)$ is given by
\begin{equation}
A[\sigma(\eta_o)] = \int_0^{2\pi} \sqrt{\tilde g_{\mu \nu} t^\mu t^\nu}~du = \int_0^{2\pi} \sqrt{\tilde g_{00} \dot \chi^2 + 2 \tilde g_{02} \dot \chi + \tilde g_{22}}~du,
\end{equation}
where $\tilde g_{\mu\nu}$ is the metric of \Eq{eq:conformallineelement} but rewritten in $(\chi,\eta_o,\theta)$ coordinates.
One finds that $\tilde g_{00} = 0$ exactly, but $\tilde g_{02}$ and $\tilde g_{22}$ do not admit any such simplifications.
Because of this, solving the full Euler-Lagrange problem to actually obtain the path $\chi(u;\eta_o)$ is intractable in general.

Nevertheless, we can exploit the fact that $\mathcal{Q}$ squeezes into the coordinate origin and perform a small-$\chi$ expansion of $A[\sigma(\eta_o)]$.
First, pull out a factor of $a(\eta_o-\chi)$ from the square root:
\begin{equation}
A[\sigma(\eta_o)] = \int_0^{2\pi} a(\eta_o-\chi) \sqrt{2f_{02}\dot \chi + f_{22}}~du
\end{equation}
In so doing we have defined $\tilde g_{\mu \nu} = [a(\eta_o - \chi)]^2f_{\mu \nu}$.
Then, expand the square root in $\chi$.
The result is
\begin{equation} \label{eq:3DasympA}
A[\sigma(\eta_o)] = \int_0^{2\pi} a(\eta_o - \chi) \left[ \frac{\chi}{R(u;\eta_o)} + \frac{1}{2} b^\prime(\eta_o) \frac{Q(u;\eta_o)}{R^2(u;\eta_o)} \chi^2 + O(\chi^3) \right]~du,
\end{equation}
where
\begin{align}
R(u;\eta_o) &= e^{-2b(\eta_o)}\cos^2 u + e^{2b(\eta_o)}\sin^2 u \nonumber \\
Q(u;\eta_o) &= e^{-2b(\eta_o)}\cos^2 u - e^{2b(\eta_o)}\sin^2 u.
\end{align}
Pulling out the scale factor is necessary to avoid pathologies that arise because both $\chi$ and $\eta_o$ become small in the same limit (see Appendix~\ref{app:example} for illustration).

Only keeping the first order term, the variation $\delta A/\delta \chi = 0$ gives
\begin{equation}
0 = -a^\prime(\eta_o-\chi) \frac{\chi}{R(u;\eta_o)} + a(\eta_o - \chi) \frac{1}{R(u;\eta_o)} \, ,
\end{equation}
so asymptotically, the maximizing path $\chi(u;\eta_o) = \chi(\eta_o)$ is given implicitly by the solution of
\begin{equation}
\chi = \frac{a(\eta_o - \chi)}{a^\prime(\eta_o-\chi)}.
\end{equation}
To first order, $A[\sigma(\eta_o)]$ is given by
\begin{equation}
A[\sigma(\eta_o)] = 2\pi \frac{a^2(\eta_o - \chi)}{a^\prime(\eta_o-\chi)} .
\end{equation}
But the requirement that $S_\mathrm{gen} \rightarrow S_\mathrm{max}$ means that $A[\sigma(\eta_o)]/4G$ must tend to the constant value $S_\mathrm{max}$, or in other words,
\begin{equation}
\lim_{\substack{\eta_o \rightarrow \eta_\infty \\ \chi \rightarrow 0}} \frac{a^2(\eta_o - \chi)}{a^\prime(\eta_o-\chi)} = \frac{2GS_\mathrm{max}}{\pi} \equiv \frac{1}{H} \, .
\end{equation}
Therefore, $a(\eta)$ asymptotically approaches de Sitter, $a(\eta) \rightarrow -1/H\eta$ as $\eta \rightarrow 0^-$, with $H = \pi/2GS_\mathrm{max}$.

Since $\chi(\eta_o)$ is a function of $\eta_o$, a technical detail to address is to check that the higher-order coefficients in the expansion \eqref{eq:3DasympA}, which themselves depend on $\eta_o$ through $b(\eta_o)$ and its derivatives, do not cause the higher-order terms to be larger than the term that is first-order in $\chi$.
This we can achieve by bounding the remainder, $r_1(\chi;\eta_o) \equiv \sqrt{2f_{02}\dot \chi + f_{22}} - \chi/R$.

Let $F = \sqrt{2f_{02}\dot \chi + f_{22}}$.
We may write its second derivative with respect to $\chi$ as
\begin{equation}
\frac{\partial^2 F}{\partial \chi^2} = b^\prime(\eta_o-\chi) \frac{Q(u; \eta_o - \chi)}{R^2(u; \eta_o-\chi)} + \varepsilon(\chi; \eta_o),
\end{equation}
where the term $\varepsilon(\chi; \eta_o) \rightarrow 0$ as $\chi \rightarrow 0$ for any $\eta_o$.
As such, choose $\chi$ and $\eta_o$ both small enough such that $|\varepsilon(\chi; \eta_o)| < |b^\prime(\eta_o-\chi)|/R(u;\eta_o-\chi)$ for all $u$.\footnote{The only instance in which this is not possible is if $|b^\prime(\eta_o-\chi)|/R(u;\eta_o-\chi)$ vanishes faster than $|\varepsilon(\chi; \eta_o)|$.
But, in this case, the remainder $|r_1(\chi;\eta_o)|$ can be bounded arbitrarily tightly, since $|\partial^2 F / \partial \chi^2|$ can be made arbitrarily small.
}
With this choice, and since $|Q/R| \leq 1$, we have that
\begin{equation}
\left| \frac{\partial^2 F}{\partial \chi^2} \right| < \frac{2 |b^\prime(\eta_o-\chi)|}{R(u; \eta_o-\chi)} \, .
\end{equation}
Next we invoke the monotonicity Assumption $(iii)$.
Let $\eta_\star = \max \{\eta_\mrm{mono}, \eta_\mrm{time} \}$.
In terms of $a(\eta)$ and $b(\eta)$, Assumption $(iii)$ reads $( a(\eta) e^{+ b(\eta)})^\prime > 0$ and $( a(\eta) e^{- b(\eta)})^\prime > 0$, or $|b^\prime(\eta)| < a^\prime(\eta)/a(\eta)$.
Therefore, upon additionally requiring $0 > \eta_o - \chi > \eta_\star$ (i.e. possibly making $\chi$ and $\eta_o$ smaller), it follows that
\begin{equation}
\left| \frac{\partial^2 F}{\partial \chi^2} \right| < \frac{2}{R(u; \eta_o-\chi)} \frac{a^\prime(\eta_o-\chi)}{a(\eta_o-\chi)} \; \longrightarrow \; \frac{2}{R(u; \eta_o-\chi)\chi(\eta_o)} \, .
\end{equation}
So, by Taylor's remainder theorem we have that
$|r_1(\chi; \eta_o)| < R(u;\eta_o-\chi_1)^{-1}(\chi^2/\chi(\eta_o))$ on any interval $\chi \in [\chi(\eta_o),\chi_2]$, where $\chi_1 \in [\chi(\eta_o),\chi_2]$ minimizes $R(u;\eta_o-\chi)$.
Or, at the edge of the interval,
\begin{equation}
|r_1(\chi(\eta_o); \eta_o)| < \frac{\chi(\eta_o)}{R(u; \eta_o-\chi_1)} \, .
\end{equation}
Since $\int_0^{2\pi} R(u; \eta_o)^{-1}~du = 2\pi$ irrespective of the value of $\eta_o$, it follows that remainder in the expansion is strictly smaller than the first-order term, so we were safe in restricting our attention to the first-order solution.

\subsubsection{Showing that the anisotropy decays.}

The decay of anisotropy is directly implied by Assumption $(iii)$ once we have established that the volumetric scale factor $a(\eta)$ is asymptotically de Sitter.
In the far future limit, Assumption $(iii)$ recast as $( a(\eta) e^{\pm b(\eta)})^\prime > 0$ gives
\begin{equation}
|b^\prime(\eta)| < \frac{a^\prime(\eta)}{a(\eta)} \; \overset{\eta \rightarrow 0^-}{\longrightarrow} \; H a(\eta) = \frac{1}{-\eta} \, .
\end{equation}
Therefore, to capture the asymptotic scaling of $b^\prime(\eta)$, we can write
\begin{equation} \label{eq:bpasymp}
b^\prime(\eta) = \frac{f(\eta)}{(-\eta)^{1-p}},
\end{equation}
where $p>0$ and where $|f(\eta)| \leq F$ for some bounded constant $F$ when $\eta > \eta_\star$.
In other words, $b^\prime(\eta)$ cannot grow faster than $1/\eta$ as $\eta \rightarrow 0^-$, so that $(-\eta)^{1-p}b^\prime(\eta)$ is some bounded function.
To establish that the anisotropy decays, and thus complete the proof of the theorem, we need only establish that $b(\eta)$ goes to a fixed limit at late times:
\begin{lem}
If $b^\prime(\eta)$ satisfies \Eq{eq:bpasymp} on $(\eta_\star,0)$, then $\lim_{\eta \rightarrow 0^-} b(\eta)$ exists.
\end{lem}
\pf We show that the limit exists by showing that $b(\eta)$ is a Cauchy function.
Let $\epsilon > 0$.
We must find $\delta > 0$ such that $0 < -\eta_1 < \delta$ and $0 < -\eta_2 < \delta$ implies that $|b(\eta_2) - b(\eta_1)| < \epsilon$.
Without loss of generality, suppose that $\eta_\star < \eta_1 < \eta_2$.
Then:
\begin{align}
\left| b(\eta_2) - b(\eta_1) \right| &= \left| \int_{\eta_1}^{\eta_2}  \frac{f(u)}{(-u)^{1-p}}~du \right| \nonumber\\
&\leq \int_{\eta_1}^{\eta_2}  \frac{|f(u)|}{(-u)^{1-p}}~du \nonumber\\
&\leq F \int_{\eta_1}^{\eta_2}  \frac{1}{(-u)^{1-p}}~du \nonumber\\
&\leq \frac{F}{p}(-\eta_1)^p.
\end{align}
Therefore, let $\delta = (\epsilon F/p)^{1/p}$.
Then $|b(\eta_2) - b(\eta_1)| < \epsilon$ as required. \eoplem

\eop

It is interesting to briefly consider how one may relax the assumption that each light cone has a globally maximal generalized entropy section.\footnote{A particularly astute reader may have noticed that the light cones in the example in \App{app:example} do not satisfy this global maximality property, but this is just because the approximation in which $S_\mrm{out}$ is estimated by $S_\mrm{CG}$ breaks down. More precisely, $S_\mrm{CG}$ is not a good estimate of the matter contribution to generalized entropy for light cone slices that are far to the past of the light cone's tip.
For such slices, the comoving volume enclosed by the slice grows arbitrarily large.}
If we do not assume that each light cone has a maximum generalized entropy surface, then the proof above pauses at \Eq{eq:SgenLowerBound}.
In this case, it is no longer true that $S_\mrm{gen}[\sigma(\tilde \eta_o),\Sigma(\tilde \eta_o)]$ must be greater than $S_\mrm{gen}[\varsigma(\tilde \eta; \tilde \eta_o),X(\tilde \eta)]$; the generalized entropy of the leaf $\sigma(\tilde \eta_o)$ could just be a local maximum, and the entropy of the constant-$\tilde \eta$ slice $\varsigma(\tilde \eta; \tilde \eta_o)$ could be larger.
We must therefore make a slightly different argument.
It turns out that a weaker but sufficient assumption is to only assume that each light cone has a unique maximum \emph{area} surface.

\begin{figure}[h]
\centering
\includegraphics[scale=0.95]{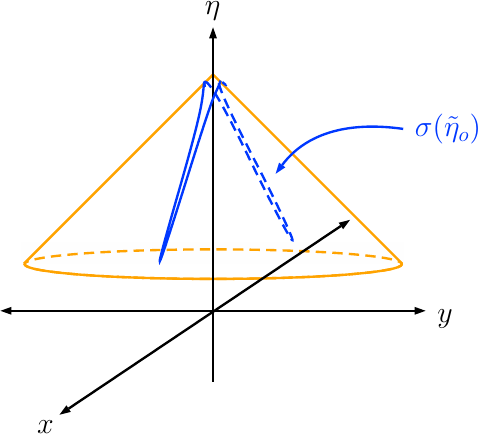}
\caption{A hypothetical leaf $\sigma(\tilde \eta_o)$ that remains bounded away from the comoving coordinate origin. The leaf has two long tendrils that extend out from the comoving coordinate origin.}
\label{fig:tendrils}
\end{figure}

As before, let us still suppose that $\mathcal{Q}$ never squeezes into the comoving coordinate origin and find a contradiction. 
We again suppose that there exists $M>0$ such that, given any $\eta_o > \eta_\mrm{time}$, one can find values $\tilde \eta_o > \eta_o$ and $\tilde u$ such that $\chi(\tilde u; \tilde \eta_o) \geq M$.
First note that in order for $S_\mrm{gen}[\sigma(\tilde \eta_o),\Sigma(\tilde \eta_o)]$ to remain finite, it must be that the function $\chi(u; \tilde \eta_o)$ is only greater than $M$ on an interval that has vanishing measure in the limit as $\tilde \eta_o \rightarrow \eta_\infty$.
Otherwise, the proper area of $\sigma(\tilde \eta_o)$ diverges.
Therefore, the leaves of $\mathcal{Q}$ develop ``tendrils'' in the asymptotic future limit, as illustrated in \Fig{fig:tendrils}.
In this case, however, the comoving volume that is enclosed by $\sigma(\tilde \eta_o)$ vanishes as $\tilde \eta_o \rightarrow \eta_\infty$, which means that the (locally) maximal entropy slice of each light cone coincides with the (globally) maximal area slice in the asymptotic future limit.
We can then repeat the same arguments presented in the proof above for the constant-$\tilde \eta$ slice, but applied in comparison to the maximal area slice, to construct the required contradiction.
Once it is established that $\mathcal{Q}$ squeezes into the comoving coordinate origin, the proof continues as before.

This relaxation is interesting (albeit somewhat artificial) because it makes it possible to avoid assuming the Quantum Focusing Conjecture.
Moreover, as is shown in \App{app:hscreenContinuity}, if a RW spacetime admits a \emph{continuous} holographic screen that has maximal area leaves on every past-directed light cone, then the screen itself is unique and there is always a finite globally maximal area slice of each light cone.
(However, this slice is not necessarily unique and may not be part of the unique continuous holographic screen with leaves on every past-directed light cone.)
This result suggests that it might in general be possible to relate continuity properties of screens to the properties of extremal-area light cone slices.
For practical purposes, however, it is much cleaner to simply assume the QFC (which also guarantees that the GSL holds).

\subsection{1+3 dimensions}

Now suppose that $\mathcal{M}$ is a Bianchi spacetime in 1+3 dimensions with the line element
\begin{equation} \label{eq:bianchiI4d}
ds^2 = -dt^2 + a_1^2(t) \, dx^2 + a_2^2(t) \, dy^2 + a_3^2(t) \, dz^2 \, .
\end{equation}
The case where $\mathcal{M}$ has 3 dimensions of space parallels the 1+2-dimensional case with only a handful of technical complications.
The main difference is that now the anisotropy has two functional degrees of freedom:
\begin{equation} \label{eq:bianchiI4d_ab}
ds^2 = -dt^2 + a^2(t)\left[e^{2b_1(t)} dx^2 + e^{2b_2(t)} dy^2 + e^{2b_3(t)} dz^2 \right]
\end{equation}
One arrives at the equation above by setting $a_i(t) = a(t)e^{b_i(t)}$ for $i = 1, 2, 3$, where the $b_i(t)$ are subject to the constraint $\sum_{i=1}^3 b_i(t) = 0$.
The definition of conformal light cone coordinates $(\eta, \eta_o, \theta, \phi)$ is correspondingly modified:
\begin{equation} \label{eq:comovingLC4D}
 x^j(\eta,\eta_o,\theta,\phi) = D^j(\theta,\phi) \int_{\eta}^{\eta_o}   \frac{e^{-2b_j(\zeta)}}{ \sqrt{ \sum_{i=1}^3 D^i(\theta,\phi)^2 \, e^{-2b_i(\zeta)} } }~d\zeta,
\end{equation}
where
\begin{equation*}
D^j(\theta,\phi) = (\sin \theta \, \cos \phi , \sin \theta \, \sin \phi , \cos \theta).
\end{equation*}
Nevertheless, the essential construction remains unchanged.
We still consider a past Q-screen, $\mathcal{Q}$, constructed with respect to a foliation of $\mathcal{M}$ by past-directed light cones, and the leaves of $\mathcal{Q}$ are still labeled by the conformal time $\eta_o$ where the tip of their corresponding light cone is located.
The no-hair theorem also generalizes in a straightforward way:
\begin{thm}
Let $\mathcal{M}$ be a Bianchi I spacetime with the line element \eqref{eq:bianchiI4d} and whose matter content has constant thermodynamic entropy $s$ per comoving volume.
Suppose that $\mathcal{M}$ admits a past Q-screen, $\mathcal{Q}$, with globally maximal entropy leaves constructed with respect to a foliation of $\mathcal{M}$ with past-directed light cones that are centered on the origin, $x = y = z = 0$.
Suppose that the Generalized Second Law holds on $\mathcal{Q}$ and that $\mathcal{M}$ and $\mathcal{Q}$ together satisfy the following assumptions for $i \in \{1,2,3\}$:
\begin{itemize}
\item[$(i)$] $a_i(t) \rightarrow \infty$ as $t \rightarrow \infty$,
\item[$(ii)$] $\mathcal{Q}$ is timelike past some $t_\mrm{time}$ and extends out to future timelike infinity,
\item[$(iii)$] $\dot a_i(t) > 0$ past some $t_\mrm{mono}$,
\item[$(iv)$] $S_\mrm{gen} \rightarrow S_\mrm{max} < \infty$ along $\mathcal{Q}$.
\end{itemize}
Then, $\mathcal{M}$ is asymptotically de Sitter and the axial scale factors $a_i(t)$ approach $C_i e^{Ht}$, where $H$ and $C_i$ are constants.
\end{thm}

\pagebreak

{\bf Note:} In terms of $a(\eta)$ and the $b_i(\eta)$, Assumption $(i)$ becomes:
\begin{itemize}
\item[$(i^\prime)$] $a(\eta) \rightarrow \infty$ as $\eta \rightarrow \eta_\infty$ and $a(\eta)e^{b_i(\eta)} \rightarrow \infty$.
\end{itemize}
\pf
The proof for 1+3 dimensions exactly parallels the proof of Theorem~\ref{thm:CNH1+2}, so we only note the most important modifications.
Beginning with Part 1, in $(\eta,\eta_o,\theta,\phi)$ coordinates, the leaves $\sigma(\eta_o)$ are now parameterized surfaces,
\begin{equation}
\tilde x^\mu(u,v; \eta_o) = (\eta(u,v; \eta_o), \eta_o, u, v) \qquad u \in [0,\pi] \qquad v \in [0,2\pi) \, .
\end{equation}
Our first task is again to show that $\chi(u,v;\eta_o) \equiv \eta_o - \eta(u,v;\eta_o)$ tends to zero for all values of $u$ and $v$ as $\eta_o \rightarrow \eta_\infty$.

As before, let us construct a contradiction of Assumption $(iv)$ by supposing that $\mathcal{Q}$ never squeezes into the comoving coordinate origin.
Suppose that there exists $M > 0$ such that, given any $\eta_o > \eta_\mrm{time}$, one can find values $\tilde \eta_o > \eta_o$, $\tilde u$, and $\tilde v$ such that $\chi(\tilde u, \tilde v; \tilde \eta_o) \geq M$.
Let $\tilde \eta \equiv \eta(\tilde u, \tilde v; \tilde \eta_o)$ and consider the constant $\eta = \tilde \eta$ slice of the light cone whose tip is at $\tilde \eta_o$.
Denote this (co-dimension 2) surface by $\varsigma(\tilde \eta ; \tilde \eta_o)$, and denote the (co-dimension 1) hypersurface of constant-$\tilde \eta$ by $X(\tilde \eta)$.
Here as well, \Eq{eq:SgenLowerBound} will lead us to the contradiction via a divergence in $A[\varsigma(\tilde \eta; \tilde \eta_o)]$.

In 1+3 dimensions, the induced metric on a surface of constant $\eta$ and $\eta_o$ is given by
\begin{equation}
\gamma_{ab} = \frac{\partial x^\mu}{\partial \theta^a} \frac{\partial x^\nu}{\partial \theta^b} g_{\mu\nu} = a^2(\eta) \sum_{j=1}^3 e^{2b_j(\eta)} \frac{\partial x^j}{\partial \theta^a} \frac{\partial x^j}{\partial \theta^b} \, ,
\end{equation}
and where $x^j$ and $g_{\mu\nu}$ refer to \Eqs{eq:bianchiI4d_ab}{eq:comovingLC4D} with $\theta^a \equiv (\theta,\phi)$.
The area of this surface is now given by the surface integral
\begin{equation} \label{eq:area4D}
A(\eta,\eta_o) = \int_0^\pi \int_0^{2\pi} \sqrt{\gamma}~d\phi \, d\theta \, ,
\end{equation}
where the determinant of the induced metric is
\begin{equation}
\gamma = a^4(\eta) \sum_{i<j} e^{2(b_i(\eta) + b_j(\eta))} \left( \frac{\partial x^i}{\partial \theta} \frac{\partial x^j}{\partial \phi} - \frac{\partial x^j}{\partial \theta} \frac{\partial x^i}{\partial \phi} \right)^2 \, .
\end{equation}
One may therefore bound the area of $\varsigma(\eta;\eta_o)$ by, e.g.,
\begin{equation}
A(\eta,\eta_o) \geq a^2(\eta) e^{b_1(\eta) + b_2(\eta)} \iint d\theta \, d\phi ~  \left| \frac{\partial x}{\partial \theta} \frac{\partial y}{\partial \phi} - \frac{\partial y}{\partial \theta} \frac{\partial x}{\partial \phi} \right| \, .
\label{eq:areaBound2D}
\end{equation}
Using the coordinate transformation \Eq{eq:comovingLC4D}, one can show that the Jacobian in the integrand above is given by
\begin{align}
\nonumber \left| \frac{\partial x}{\partial \theta} \frac{\partial y}{\partial \phi} - \frac{\partial y}{\partial \theta} \frac{\partial x}{\partial \phi} \right| = \iint_\eta^{\eta_o} ds \, ds^\prime ~ & \sin \theta \left| \cos \theta \right| \left( \sin^2 \theta \cos^2 \phi \, e^{2(b_2(s)+b_3(s^\prime))} \right. \\
\nonumber & \left. + \sin^2 \theta \sin^2 \phi \, e^{2(b_3(s)+b_1(s^\prime))} + \cos^2 \theta \, e^{2(b_2(s)+b_1(s^\prime))} \right) \\
& \times \left( \sum_{i=1}^3 D^i(\theta,\phi)^2 \, e^{-2b_i(s)} \right)^{-3/2} \left( \sum_{j=1}^3 D^j(\theta,\phi)^2 \, e^{-2b_i(s^\prime)} \right)^{-3/2} \, .
\label{eq:beast1}
\end{align}
This is quite beastly, but fortunately we can bound it nicely:
\begin{align}
\nonumber \left| \frac{\partial x}{\partial \theta} \frac{\partial y}{\partial \phi} - \frac{\partial y}{\partial \theta} \frac{\partial x}{\partial \phi} \right| \geq \iint_\eta^{\eta_o} ds \, ds^\prime ~ & \sin \theta \left| \cos^3 \theta \right| e^{2(b_2(s)+b_3(s^\prime))} \\
\nonumber  & \times \left((e^{-2b_1(s)} + e^{-2b_2(s)}) \sin^2\theta + e^{-2b_3(s)} \cos^2 \theta \right)^{-3/2} \\
 & \times \left((e^{-2b_1(s^\prime)} + e^{-2b_2(s^\prime)}) \sin^2\theta + e^{-2b_3(s^\prime)} \cos^2 \theta \right)^{-3/2}
 \label{eq:beast2}
\end{align}
(One would arrive at similar results by choosing different terms to keep in the numerator of \Eq{eq:beast1}.)
Then, inserting \Eq{eq:beast2} into \Eq{eq:areaBound2D} and performing the angular integration, one arrives at
\begin{align}
\nonumber A(\eta,\eta_o) & \geq 4 \pi a^2(\eta) e^{b_1(\eta)+b_2(\eta)} \iint_\eta^{\eta_o} ds \, ds^\prime ~ \tilde f(s,s^\prime)
\end{align}
where
\begin{align*}
\tilde f(s,s^\prime) &= \frac{e^{b_1(s) + 3b_2(s) + 3b_1(s^\prime) + b_2(s^\prime)}}{\left[ \sqrt{e^{2b_1(s)} + e^{2b_2(s)}} e^{2(b_1(s^\prime) + b_2(s^\prime))}  + \sqrt{e^{2b_1(s^\prime)} + e^{2b_2(s^\prime)}} e^{2(b_1(s) + b_2(s))} \right]^2} \\
&\geq \frac{e^{b_1(s) + 3b_2(s) + 3b_1(s^\prime) + b_2(s^\prime)}}{ \left[ (e^{b_1(s)} + e^{b_2(s)}) e^{2(b_1(s^\prime) + b_2(s^\prime))}  + (e^{b_1(s^\prime)} + e^{b_2(s^\prime)}) e^{2(b_1(s) + b_2(s))} \right]^2  } \\
& \equiv f(s,s^\prime) \, .
\end{align*}
Note that we have also used the fact that $b_3 = -b_1 - b_2$ to eliminate $b_3$.
Then, if $f(s,s^\prime)$ is minimized at $s_m, s^\prime_m \in [\eta,\eta_o]$, it follows that
\begin{equation}
A(\eta,\eta_o) \geq 4 \pi (\eta_o - \eta)^2 a^2(\eta) e^{b_1(\eta)+b_2(\eta)} f(s_m,s^\prime_m) \, .
\end{equation}

Given this result, we now apply it to our surface $\varsigma(\tilde \eta; \tilde \eta_o)$, for which $(\tilde \eta_o - \tilde \eta) \geq M$.
Doing so, we arrive at
\begin{equation} \label{eq:4Dbound}
A[\varsigma(\tilde \eta; \tilde \eta_o)] \equiv A(\tilde \eta, \tilde \eta_o) \geq 4 \pi M^2 a^2(\tilde \eta) e^{b_1(\tilde \eta)+b_2(\tilde \eta)} f(s_m,s^\prime_m) \, .
\end{equation}
The right-hand side of the bound above then diverges as $\tilde \eta$ and $\tilde \eta_o$ are chosen arbitrarily large.
The only subtlety arises if either or both of $b_1$ and $b_2$ also diverge, but because the numerator and the denominator of the $b$-dependent part of the bound \Eq{eq:4Dbound} contain equal powers of $b_1$ and $b_2$, the overall divergent behaviour induced by $a(\eta)$ is unchanged.
(Recall that $a(\eta) e^{b_1(\eta)}$, $a(\eta) e^{b_2(\eta)}$, and $a(\eta) e^{b_3(\eta)} = a(\eta) e^{-b_1(\eta) - b_2(\eta)}$ all grow infinitely large by assumption.)
We therefore arrive at the desired contradiction of Assumption $(iv)$ via \Eq{eq:SgenLowerBound}, and so the leaves of $\mathcal{Q}$ squeeze into the comoving coordinate origin in the asymptotic future.

Next we turn to showing that the scale factor $a(\eta)$ is asymptotically de Sitter (Part 2).
Consider the generalized entropy $S_\mrm{gen}[\sigma(\eta_o),\Sigma(\eta_o)]$ once more.
First, Lemma~\ref{lem:constSlice} is correspondingly modified:

\begin{lem} \label{lem:constSlice4D}
Let $\varsigma(\eta; \eta+\chi)$ be the constant-$\eta$ slice of the past-directed light cone whose tip is at $\eta_o = \eta + \chi$.
The generalized entropy defined by this slice is given by
\begin{equation}
S_\mathrm{gen}[\varsigma(\eta;\eta+\chi),X(\eta)] = \frac{A(\eta,\eta+\chi)}{4G} + c_g(\eta,\chi) \chi^3 s,
\end{equation}
where $A(\eta,\eta+\chi)$ is given by
\begin{equation} \label{eq:A4Dasymptotic}
A(\eta,\eta+\chi) = a^2(\eta) \cdot \left[ 4\pi \chi^2 + O(\chi^4) \right],
\end{equation}
and $c_g(\eta,\chi)$ is some $O(1)$ geometric factor due to anisotropy that does not depend on $a(\eta)$.
\end{lem}

\pf
Repeating the steps described in Lemma~\ref{lem:constSlice}, one finds that
\begin{equation}
c_g(\eta,\chi) \equiv \frac{S_\mathrm{CG}(\eta,\chi)}{\chi^3 s} = \frac{4\pi}{3} + \frac{8\pi}{45} \left(b_1^\prime(\eta)^2 + b_1^\prime(\eta) b_2^\prime (\eta) + b_2^\prime(\eta)^2 \right) \chi^2 + O(\chi^3).
\end{equation}

The expansion of $A(\eta,\eta+\chi)$ for small $\chi$ follows from expanding $\sqrt{\gamma}$ in \Eq{eq:area4D} in powers of $\chi$ and then integrating.
\eoplem

~

From Lemma~\ref{lem:constSlice4D}, it therefore again follows that the matter contribution to the generalized entropy, $S_\mrm{CG}[\sigma(\eta_o),\Sigma(\eta_o)]$, vanishes in the asymptotic future limit.
Consequently, we focus on the area term, $A[\sigma(\eta_o)]$.

For this part of the proof, we will work in the coordinates $(\chi,\eta_o,\theta,\phi)$.
The leaf $\sigma(\eta_o)$ is parameterized by some surface $\tilde x^\mu(u,v) = (\chi(u,v;\eta_o),\eta_o,u,v)$ with $\eta_o$ held constant and $0 \leq u \leq \pi$, $0 \leq v < 2\pi$.
In the asymptotic future, this surface is the surface on the light cone with tip at $\eta_o$ with maximal area, and so it is the solution of
\begin{equation}
\frac{\delta A[\sigma(\eta_o)]}{\delta \chi(u,v;\eta_o)} = 0 \, .
\end{equation}
The induced metric on this surface is, as usual, given by
\begin{equation}
h_{ab} = \frac{\partial \tilde x^\mu}{\partial u^a} \frac{\partial \tilde x^\nu}{\partial u^b} \tilde g_{\mu \nu}
\end{equation}
where $\tilde g_{\mu\nu}$ is the metric of \Eq{eq:bianchiI4d_ab} but rewritten in $(\chi,\eta_o,\theta,\phi)$ coordinates.
The area of $\sigma(\eta_o)$ is given by
\begin{equation}
A[\sigma(\eta_o)] = \int_0^\pi \int_0^{2\pi} \sqrt{\det h}~dv \, du \, ,
\end{equation}
and the components of $h_{ab}$ are as follows:
\begin{align}
h_{uu} &= (\partial_u \chi)^2 \tilde g_{00} + 2(\partial_u \chi) \tilde g_{02} + \tilde g_{22} \nonumber\\
h_{uv} &= (\partial_u \chi)(\partial_v \chi) \tilde g_{00} + (\partial_u \chi) \tilde g_{03} + (\partial_v \chi) \tilde g_{02} + \tilde g_{23} \nonumber\\
h_{vv} &= (\partial_v \chi)^2 \tilde g_{00} + 2(\partial_v \chi) \tilde g_{03} + \tilde g_{33}.
\end{align}

Once more, solving the full Euler-Lagrange problem for $\chi(u,v;\eta_o)$ to obtain the maximal area $A$ is intractable, so we use the same trick where we extract an overall factor of $a^4(\eta_o-\chi)$ from $\det h$ and then expand the square root of the quotient in powers of $\chi$.
The result is
\begin{equation}
A[\sigma(\eta_o)] = \int_0^\pi \int_0^{2\pi} a^2(\eta_o - \chi) \left[ \frac{ \sin \theta}{R(u,v;\eta_o)^{3/2}}\chi^2 + \frac{Q(u,v;\eta_o) \sin \theta}{R(u,v;\eta_o)^{5/2}} \chi^3 + O(\chi^4) \right] ~ dv\, du,
\end{equation}
where
\begin{align}
R(u,v;\eta_o) &= \sum_{i=1}^3 e^{-2b_i(\eta_o)} D^i(u,v)^2 \nonumber \\
Q(u,v;\eta_o) &= \sum_{i=1}^3 b_i^\prime(\eta_o) e^{-2b_i(\eta_o)} D^i(u,v)^2.
\end{align}

Only keeping the lowest order term, the variation $\delta A/\delta \chi = 0$ gives the maximal path $\chi(u,v;\eta_o) = \chi(\eta_o)$ as the solution of
\begin{equation}
\chi = \frac{a(\eta_o - \chi)}{a^\prime(\eta_o-\chi)}.
\end{equation}
So, to lowest order, $A[\sigma(\eta_o)]$ is given by
\begin{equation}
A[\sigma(\eta_o)] = \chi^2 a^2(\eta_o - \chi) \int_0^\pi \int_0^{2\pi} \frac{\sin \theta}{R(u,v;\eta_o)^{3/2}} ~ dv \, du = 4\pi \left(\frac{a^2(\eta_o - \chi)}{a^\prime(\eta_o-\chi)}\right)^2 \, .
\end{equation}
But the requirement that $S_\mathrm{gen} \rightarrow S_\mathrm{max}$ means that $A[\sigma(\eta_o)]/4G$ must tend to the constant value $S_\mathrm{max}$, or in other words,
\begin{equation}
\lim_{\substack{\eta_o \rightarrow \eta_\infty \\ \chi \rightarrow 0}} \frac{a^2(\eta_o - \chi)}{a^\prime(\eta_o-\chi)} = \sqrt{\frac{GS_\mathrm{max}}{\pi}} \equiv \frac{1}{H}
\end{equation}
Therefore, $a(\eta)$ asymptotically approaches de Sitter, $a(\eta) \rightarrow -1/H\eta$ as $\eta \rightarrow 0^-$, with $H^2 = \pi/GS_\mathrm{max}$.
Note that we recover the same Hubble constant as in Theorem~\ref{thm:CNH-RW} for RW spacetimes in 1+3 dimensions.

Finally, as in the case of 1+2 dimensions, the condition $(a(\eta)e^{b_i(\eta)})^\prime > 0$ is enough to show that $\lim_{\eta \rightarrow 0^-} b_i^\prime(\eta)$ exists for each $i$.
\eop

\section{Discussion}
\label{sec:disc}

Assuming the Generalized Second Law, we have shown that if a Bianchi I spacetime admits a past Q-screen along which generalized entropy increases up to a finite maximum, then this implies that the spacetime is asymptotically de Sitter.
We recover a version of Wald's cosmic no-hair theorem by making thermodynamic arguments about spacetime, without appealing to Einstein's equations.

While the proof of these cosmic no-hair theorems is most tractable (and certainly easiest to visualize) in 1+2 dimensions, the generalization to 1+3 dimensions was fairly immediate.
In principle, the proof strategy for arbitrary dimensions is the same, albeit more difficult from the perspective of calculation.
This is chiefly because calculating area elements of codimension-2 surfaces in arbitrary dimensions is cumbersome.
Nevertheless, it is natural to expect that analogous cosmic no-hair theorems hold for Bianchi I spacetimes of arbitrary dimensions.

Within the proof itself, it would be interesting to see if the monotonicity assumptions, $a^\prime_i(\eta) > 0$, could be eliminated.
The fact that the Generalized Second Law asserts that $S_\mrm{gen}$ increases monotonically along a Q-screen does offer some leverage.
In particular, asymptotically this implies that the average scale factor $a(\eta) = (\prod_{i=1}^d a_i(\eta))^{1/d}$ increases monotonically; however, we learn nothing about the anisotropies $b_i(\eta)$, since the leading order behaviour of $S_\mrm{gen}$ does not depend on the $b_i(\eta)$ in the asymptotic future regime.
We also note that the monotonicity assumptions do not trivialize the cosmic no-hair theorems demonstrated in \Sec{sec:Bianchi}.
For example, assuming monotonicity does not rule out exponential expansion with different rates in different spatial directions, nor asymptotically power-law scale factors, nor does it even imply accelerated expansion at all.

An interesting extension would be to try to prove a no-hair theorem for classical cosmological perturbations \cite{Mukhanov1992}, or for quantum fields in curved spacetime.
Given a scalar field on a curved spacetime background, the task is to show that the combined metric and scalar field perturbations approach the Bunch-Davies state \cite{Birrell1982} at late times.
In principle it would suffice to show that the background spacetime still tends to de Sitter in the future in this case, since one could then simply invoke known no-hair results about scalar fields in curved backgrounds \cite{Hollands2010,Marolf2010,Marolf2011}.
Conceptually, such a calculation would be interesting because one can explicitly write down the quantum state of cosmological perturbations, and so a full treatment of the matter entropy as von Neumann entropy (modulo ultraviolet divergences) is possible.

To prove our theorem, it was not strictly necessary to assume that the gravitational contribution to the entropy was precisely proportional to the surface area.
We could imagine choosing some other function of the area, such that
\begin{equation}
S_\mrm{gen}[\sigma,\Sigma] = f(A[\sigma]/G) + S_\mrm{CG}[\sigma,\Sigma].
\end{equation}
For example, returning to the RW case, if one sets $f(A/G) = C(A/G)^p$ for some constants $C$ and $p$, exactly the same analysis as in the proof of Theorem~\ref{thm:CNH-RW} leads to the conclusion that (cf. \Eq{eq:RWasymp})
\begin{equation}
\dot a(t) \rightarrow \sqrt{\frac{4\pi}{G} \left(\frac{C}{S_\mrm{max}}\right)^{1/p}} a(t)
\end{equation}
in the limit as $t \rightarrow \infty$.
In other words, one still concludes that the scale factor is asymptotically de Sitter, albeit with a Hubble constant that differs from the usual case of $f(A/G) = A/4G$.

Finally, while we did not make use of the Einstein field equations in our derivation, upon reinvoking them, we note that the cosmic no-hair theorems established here imply a pure dark energy phase asymptotically in the future (in the sense that the stress energy tensor becomes proportional to the metric, $g_{\mu\nu}$).
However, the GSL is not sensitive to the nature of the dark energy (whether it is a pure cosmological constant, whether it turns on, whether it's due to a slowing scalar field, and so on).

This work can be thought of as part of the more general program of connecting gravitation to entropy, thermodynamics, and entanglement \cite{jacobson95,Padmanabhan:2009vy,Verlinde,jacobson15,Carroll:2016lku,Swingle:2009,VanRaamsdonk,Lashkari,Faulkner,Swingle:2012,graventanglement2,er-eprms,Cao:2016mst}.
As in attempts to derive Einstein gravity from entropic considerations, we deduce the behavior of the geometry of spacetime from thermodynamics, without explicit field equations.
Our result is less general, as we only obtain the asymptotic behavior of the universe, but is perhaps also more robust, as our assumptions are correspondingly minimal.
Thinking of spacetime as emerging thermodynamically from a set of underlying degrees of freedom can change our perspective on the knotty problems of quantum gravity; for example, as emphasized by Banks \cite{Banks:2000fe}, the cosmological constant problem becomes the question of ``Why does Hilbert space have a certain number of dimensions?" rather than ``Why is this parameter in the low-energy effective Lagrangian so small?"
Problems certainly remain (including why the entropy was so low near the Big Bang), but this alternative way of thinking about gravitation may prove useful going forward.

\section*{Acknowledgments}

We would like to thank Cliff Cheung, John Preskill, and Alan Weinstein for helpful discussions.
This material is based upon work supported by the U.S. Department of Energy, Office of Science, Office of High Energy Physics, under Award No. DE-SC0011632, as well as by the Walter Burke Institute for Theoretical Physics at Caltech and the Gordon and Betty Moore Foundation through Grant No. 776 to the Caltech Moore Center for Theoretical Cosmology and Physics.

\appendix

\section{Q-screens, a worked example}
\label{app:example}

In this appendix, we illustrate Q-screens by explicitly constructing one in a RW spacetime that is asymptotically de Sitter.
Consider a RW spacetime in 1+3 dimensions with the line element $ds^2 = -dt^2 + a^2(t)(d\chi^2 + \chi^2 d\Omega_2^2)$ and where the scale factor is $a(t) = \sinh t$, $t \in (0,\infty)$.
Conformal time is given by $\eta(t) = -2 \, \mrm{arccoth}(e^t)$, $\eta \in (-\infty,0)$, and the scale factor in conformal time is
\begin{equation}
a(\eta) = \frac{1}{\sinh(-\eta)} \, .
\end{equation}
Foliate the spacetime with past-directed light cones centered at the coordinate origin $\chi = 0$, and let the Cauchy hypersurfaces of the spacetime be the constant-$\eta$ hypersurfaces.
Let us now construct a Q-screen by extremizing the generalized entropy on each light cone.

\begin{figure}[h]
\includegraphics[scale=0.4]{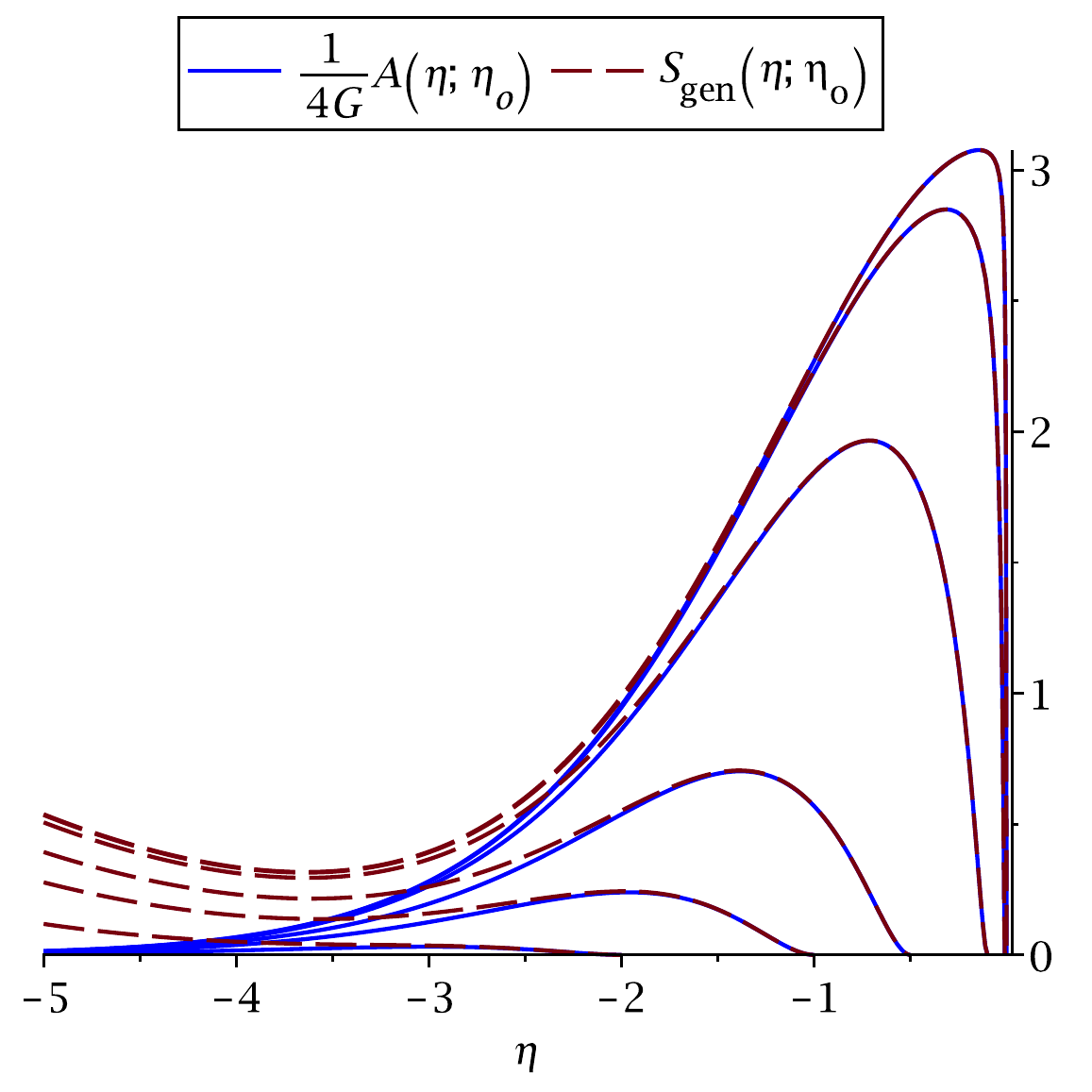}
\caption{Plots of area (solid) and generalized entropy (dashed) along light cones.
From the lowest peak to the highest peak, the values of $\eta_o$ are $-2$, $-1$, $-0.5$, $-0.1$, $-0.01$, and $-0.001$.
Here we have taken $G=1$ and we have picked $s = 0.001$.}
\label{fig:SgenEx}
\end{figure}

Consider a past-directed light cone whose tip is at the conformal time $\eta_o$.
A constant-$\eta < \eta_o$ slice of this light cone is a 2-sphere of coordinate radius $\eta_o - \eta$, and so the generalized entropy computed with respect to this slice is
\begin{equation}
S_\mrm{gen}(\eta;\eta_o) = \frac{\pi}{G} \left(\frac{\eta_o - \eta} {\sinh(-\eta)}\right)^2 + \frac{4}{3} \pi (\eta_o - \eta)^3 s \, .
\end{equation}
A plot of $S_\mrm{gen}(\eta;\eta_o)$ as a function of $\eta$ for several values of $\eta_o$ is shown in \Fig{fig:SgenEx}.
The area term $A(\eta;\eta_o)/4G$ alone is also overlaid on the plot, which illustrates that it is the dominant contribution to the generalized entropy at late times.
Notice that in addition to having a local maximum, $S_\mrm{gen}(\eta;\eta_o)$ also has a local minimum, and below a certain critical value $\eta_o^\mrm{crit}$ there is in fact no nonzero value of $\eta$ which locally extremizes $S_\mrm{gen}(\eta;\eta_o)$.
As such, the Q-screen, which is defined as the union of the slices with maximal generalized entropy, is only defined for $\eta_o \geq \eta_o^\mrm{crit}$.
This is in contrast to the area $A(\eta; \eta_o)$, which has a locally maximizing value of $\eta$ for all $\eta_o$.
The holographic screen, which is made up of extremal \emph{area} slices, is therefore defined for all times.
Both the Q-screen and the holographic screen were schematically illustrated previously in \Fig{fig:QscreenEx}.

\begin{figure}[h]
\includegraphics[scale=0.4]{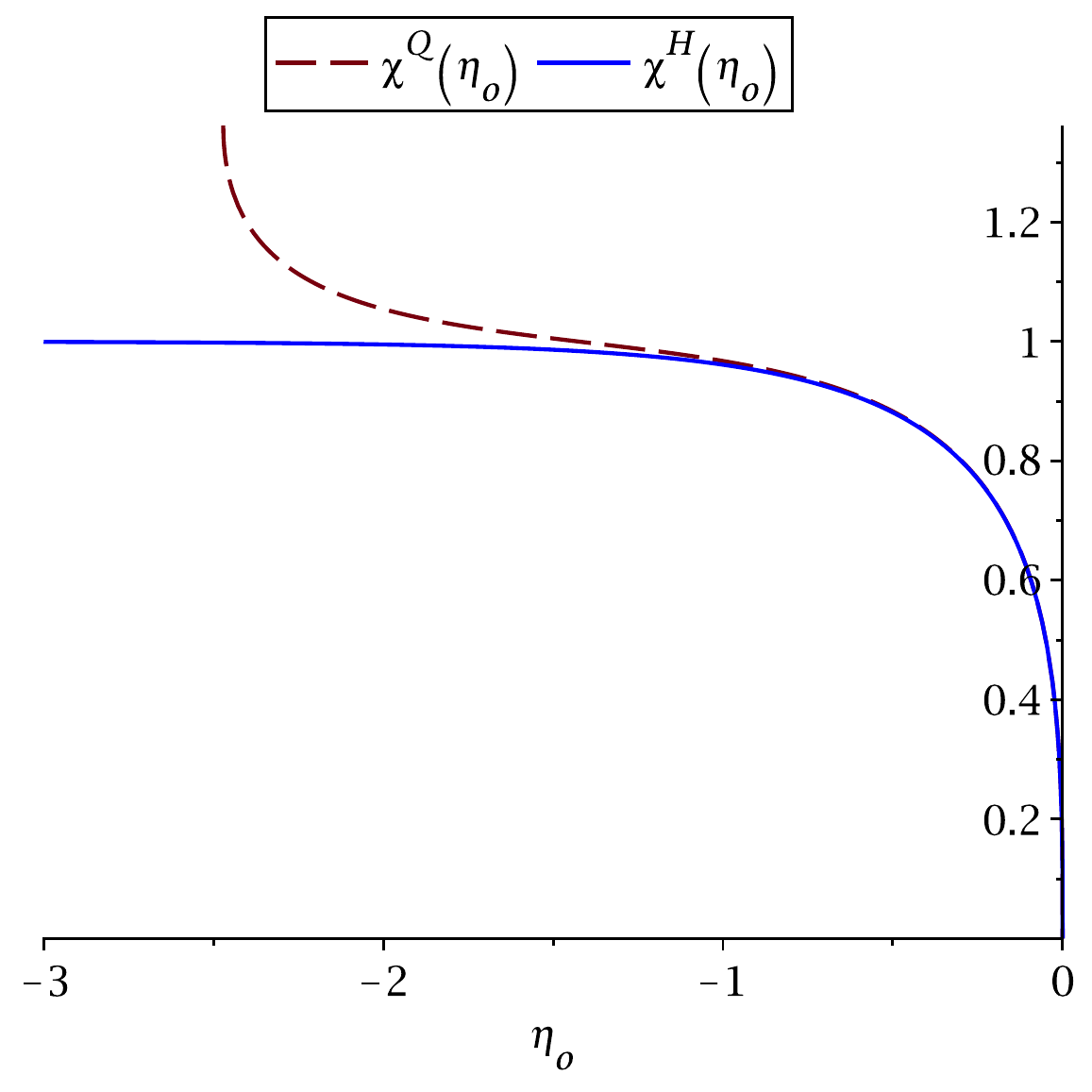}
\caption{Asymptotic behaviour of the radius of the Q-screen leaves ($\chi^ Q(\eta_o)$, dashed) and holographic screen leaves ($\chi^ H(\eta_o)$, solid).}
\label{fig:chimax}
\end{figure}

Generalized entropy is extremal when $\partial S_\mrm{gen}/\partial \eta = 0$.
Excluding $\eta = 0$ and $\eta \rightarrow -\infty$, the extremizing values of $\eta$ are the real-valued solutions of
\begin{equation}
\eta_o - \eta = \frac{\sinh(-\eta)}{\cosh(-\eta) - 2Gs \sinh(-\eta)^3}
\end{equation}
when they exist.
Let $\eta^Q(\eta_o)$ denote the maximizing value, and hence also define the Q-screen leaf radius $\chi^Q(\eta_o) \equiv \eta_o - \eta^Q(\eta_o)$.
A plot of $\chi^Q(\eta_o)$ is shown in \Fig{fig:chimax}.
As expected, $\chi^Q(\eta_o)$ vanishes as $\eta_o \rightarrow 0^-$.
For comparison, we also plot the holographic screen radius $\chi^H(\eta_o) \equiv \eta_o - \eta^H(\eta_o)$, where $\eta^H(\eta_o)$ maximizes the area of the light cone slice, i.e., it is the solution of
\begin{equation} \label{eq:maxAcond}
\eta_o - \eta = \tanh(-\eta) \, .
\end{equation}
In particular note that $\chi^Q(\eta_o)$ is always slightly larger than $\chi^H(\eta_o)$, but they ultimately coincide in the limit $\eta \rightarrow 0^-$ (cf. \Fig{fig:QscreenEx}).

As a final exercise, let us investigate the asymptotic dependence of $\chi^H(\eta_o)$ on $\eta_o$ (which is also the asymptotic dependence of $\chi^Q(\eta_o)$, since the two coincide as $\eta_o \rightarrow 0^-$) to illustrate some of the subtleties involved in performing asymptotic expansions.
Consider \Eq{eq:maxAcond} and let $\eta = \eta_o - \chi$ so that we have $\chi = \tanh(\chi-\eta_o)$.
Since, asymptotically, $\chi \rightarrow 0$, one may be tempted to expand this last equation for small values of $\chi$:
\begin{equation}
\chi = \tanh(-\eta_o) + (1 - \tanh^2(-\eta_o))\chi + O(\chi^2) \quad \Rightarrow \quad \chi^H(\eta_o) \overset{?}{=} \frac{1}{\tanh(-\eta_o)}
\end{equation}
Notice, however, that since $0 < \tanh(-\eta_o) < 1$, this expression for $\chi^H(\eta_o)$ cannot be infinitesimally small---the expansion is inconsistent!
Rather, $\chi$ and $\eta_o$ are simultaneously infinitesimal.
Consider instead the double Taylor series in $\chi$ and $\eta_o$:
\begin{equation}
\chi = \chi - \eta_o - \tfrac{1}{3}\chi^3 + \eta_o \chi^2 - \eta_o^2 \chi + \tfrac{1}{3} \eta_o^3 + \cdots \quad \Rightarrow \quad \chi^H(\eta_o) = (-3\eta_o)^{1/3} + \eta_o  + \cdots
\end{equation}
This last result is the correct asymptotic behaviour of $\chi^H(\eta_o)$.

Similarly, writing $A = 4\pi \chi^2 a^2(\eta_o - \chi)$, one arrives at the wrong expressions for extremal values if one tries to expand $A$ in small values of $\chi$, $\eta_o$, or even both at the same time.
The key is to keep $a(\eta_o - \chi)$ intact so that one arrives at \Eq{eq:maxAcond}.
Doing so leaves just enough nonlinearity to be able to restore the correct asymptotic behaviour of $\chi^H(\eta_o)$.
This technique is exploited in \Sec{sec:asympdS}.

\section{Holographic screen continuity and maximal area light cone slices}
\label{app:hscreenContinuity}

When the null curvature condition holds, the Raychaudhuri equation guarantees that light rays focus, or in other words, that the expansion of a null congruence is always nonincreasing: $d\theta/d\lambda \leq 0$.
In particular, this means that if a null congruence has a spacelike slice whose area is maximal with respect to local deformations, then this is in fact the unique globally maximal area slice.
A consequence of this observation is that if one's aim is to construct a holographic screen by stitching together maximal area slices of each null sheet in a null foliation, then the holographic screen is uniquely fixed by the choice of foliation.

Here, we connect the uniqueness of locally maximal area slices to continuity properties of holographic screens in RW spacetimes.
What we will first show is that, given a foliation of a RW spacetime by past-directed light cones, there is at most one continuous holographic screen that can be constructed with respect to this foliation that has maximal area leaves on every light cone.
We will then show that a consequence of this observation is that if a spacetime admits a continuous holographic with maximal area leaves on every light cone, then each light cone necessarily has a globally maximal finite area slice.

\begin{prop} \label{RWcont1}
Let $\mathcal{M}$ be a RW spacetime with the line element
\begin{equation}
ds^2 = a^2(\eta) \left( -d\eta^2 + d\chi^2 + \chi^2 d\Omega_{d-1}^2 \right),
\end{equation}
where the conformal time $\eta$ takes values in an unbounded (connected) interval $\mathcal{I} \subseteq \mathbb{R}$.
Consider a foliation of $\mathcal{M}$ by past-directed light cones whose tips are at $\chi = 0$.
If there is a past-directed light cone that has multiple spacelike slices that have maximal area with respect to local deformations, then $\mathcal{M}$ admits at most one holographic screen, $H$, constructed with respect to the given foliation that is both $(a)$ continuous, and $(b)$ has maximal area leaves on every past-directed light cone.
\end{prop}

\pf
Consider a past-directed light cone whose tip is at $\eta_o$.
For $\eta < \eta_o$, the area of the constant-$\eta$ slice of this light cone is given by
\begin{equation}
A(\eta,\eta_o) = \mathcal{N}_d \left[(\eta_o - \eta) a(\eta)\right]^{d-1},
\end{equation}
where $\mathcal{N}_d$ is a dimension-dependent constant.
Because $\mathcal{M}$ is spherically symmetric, such a slice has extremal area if $\partial A/\partial \eta = 0$, or equivalently, if
\begin{equation}
\eta_o = \eta + \frac{a(\eta)}{a^\prime(\eta)} \equiv f(\eta) \, .
\end{equation}
Therefore, constant-$\eta$ slices of the past-directed light cone whose tip is at $\eta_o$ for which $f(\eta) = \eta_o$ and $\eta < \eta_o$ are potential holographic screen leaves.

\begin{figure}[h]
\begin{minipage}[t]{0.45\textwidth}
	\centering
	\includegraphics[scale=1]{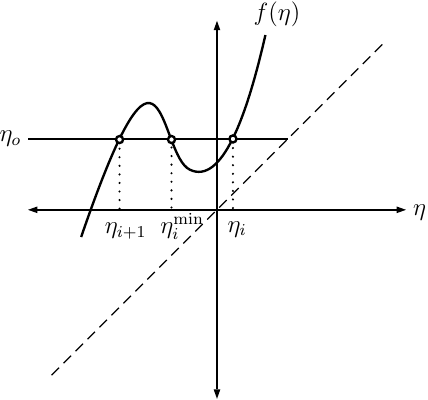}\\(a)
\end{minipage}
~~
\begin{minipage}[t]{0.45\textwidth}
	\centering
	\includegraphics[scale=1]{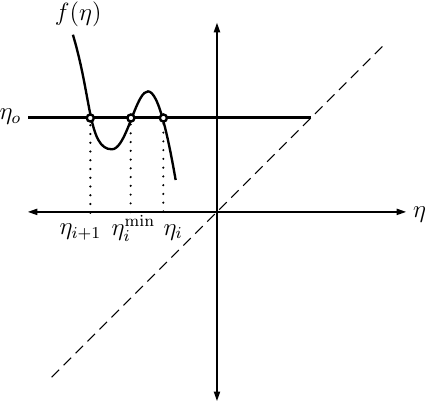}\\(b)
\end{minipage}
\\
\begin{minipage}[t]{0.45\textwidth}
	\centering
	\includegraphics[scale=1]{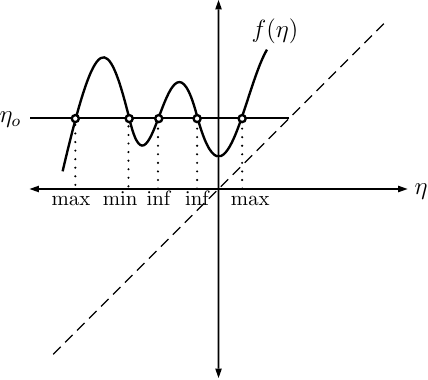}\\(c)
\end{minipage}
~~
\begin{minipage}[t]{0.45\textwidth}
	\centering
	\includegraphics[scale=1]{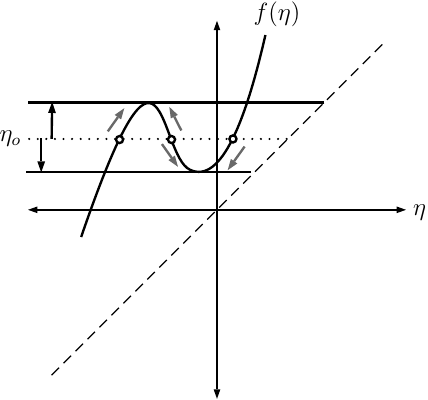}\\(d)
\end{minipage}
\caption{In the vicinity of two local maxima of $A(\eta,\eta_o)$ at $\eta_i$ and $\eta_{i+1}$ with $\eta_o$ held fixed, the graph of the function $f(\eta)$ must schematically resemble one of the configurations shown in (a) and (b). The dashed line is the $\eta_o = \eta$ line; $\eta$ can only take values to the left of this line.  The graph (c) depicts a configuration with additional inflection points. The graph (d) illustrates how the maxima and the minimum eventually meet and annihilate as the horizontal constant $\eta_o$ line is shifted up and down.}
\label{fig:hscreenf}
\end{figure}

Now suppose that there is a light cone whose tip is at $\eta_o$ that has $n$ locally maximal area slices at $\eta = \eta_1, \eta_2, ..., \eta_n$ where, for convenience, these conformal times are ordered such that $\eta_1 > \eta_2 >  \dots > \eta_n$.
This means that a graph of $f(\eta)$ must intersect the horizontal line at $\eta_o$ at least $2n-1$ times. (Between any two adjacent local maxima $\eta_i$ and $\eta_{i+1}$, there must be a local minimum of area at some $\eta^{\mrm{min}}_i \in (\eta_i,\eta_{i+1})$, and there may also be inflection points.)
Consider any two adjacent local maxima $\eta_i$ and $\eta_{i+1}$.
Schematically, in the vicinity of these points, the graph of $f(\eta)$ must look like one of the two configurations shown in \Fig{fig:hscreenf}~(a,b), since $f$ is continuous if $a/a^\prime$ is continuous.
Now consider shifting the horizontal line at $\eta_o$ up and down.
This corresponds to shifting the tip of the light cone to the future and past of $\eta_o$.
Where the horizontal line intersects the graph of $f(\eta)$ tracks how the location of the local maxima and the local minimum of $A$ move.
In particular, notice that by moving the horizontal line sufficiently far to the future or the past, one of the local maxima and the local minimum must eventually meet and become an inflection point before disappearing altogether.
(Note that we may always move the horizontal line sufficiently far in at least one of the past or future directions, since the interval $\mathcal{I}$ in which the conformal time takes its values is unbounded in at least one direction.)
Therefore, if we track how the locations of the maxima at $\eta_i$ and $\eta_{i+1}$ change as we move the location of the light cone's tip, we see that one of these local maxima must eventually disappear, as illustrated in \Fig{fig:hscreenf}~(d).

Inductively, then, there exists at most one \emph{continuous} function, call it $\eta_\mrm{max}(\eta_o)$, whose domain is all $\eta_o \in \mathcal{I}$ and is such that $\eta = \eta_\mrm{max}(\eta_o)$ is a local maximum of $A(\eta,\eta_o)$ for all $\eta_o$.
The union of the constant-$\eta_\mrm{max}(\eta_o)$ slices of all past-directed light cones is precisely the holographic screen $H$ described in the statement of the proposition.
\eop

~

\begin{figure}[h]
\begin{minipage}[t]{0.45\textwidth}
	\centering
	\includegraphics[scale=1]{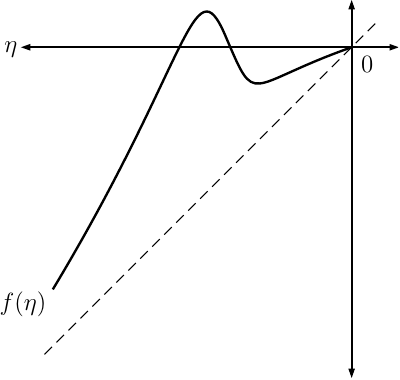}\\(a)
\end{minipage}
~~
\begin{minipage}[t]{0.45\textwidth}
	\centering
	\includegraphics[scale=1]{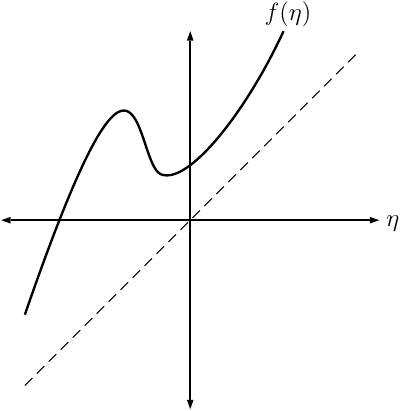}\\(b)
\end{minipage}
\caption{(a) An example of a function $f(\eta)$, where $\mathcal{I} = (-\infty, 0)$, which admits a continuous holographic screen with leaves on every light cone, but where some light cones have two local maxima of $A(\eta,\eta_o)$ when their tips are close enough to the endpoint $\eta_o = 0$. (b) An example of a function $f(\eta)$, where $\mathcal{I} = \mathbb{R}$, which admits no continuous holographic screen with leaves on every light cone.}
\label{fig:fex}
\end{figure}

Examples of various $f(\eta)$ are sketched below.
\Fig{fig:fex}~(a) depicts a case in which there exists a continuous holographic screen with leaves on every light cone.
\Fig{fig:fex}~(b) depicts a case in which there is no such holographic screen.
In fact, from this example, one can see that if $\mathcal{I} = \mathbb{R}$, then there can never be a continuous holographic screen with leaves on every light cone if there is a light cone that has multiple maximal area slices.
Referring to the proof above, the technical reason is that in this case, the horizontal line of constant $\eta_o$ can be pushed arbitrarily far up and down since $\eta_o$ can take all values in $\mathbb{R}$, and so any pair of adjacent maxima and minima will eventually merge (as a function of $\eta_o$).

Finally, there is a partial converse of the result above:

\begin{prop}
If $\mathcal{M}$ as described in Proposition~\ref{RWcont1} admits a continuous holographic screen, $H$, with maximal area leaves on every past-directed light cone, then each light cone has a spatial slice which is a global maximum of the area of all spatial slices of the light cone, and the area of this slice is finite.
\end{prop}

\begin{figure}[h]
\centering
\includegraphics[scale=1]{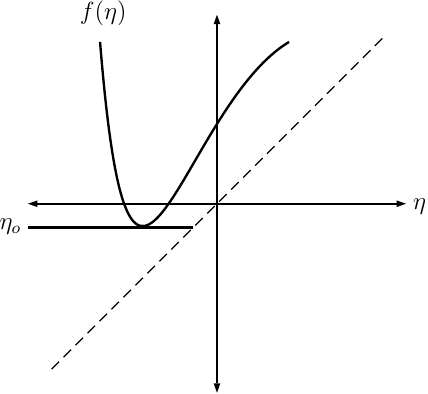}
\caption{An example of $f(\eta)$ that is concave up, with $\mathcal{I} = \mathbb{R}$. Above the horizontal $\eta_o$ line shown, this $f(\eta)$ intersects a horizontal line twice, which means that light cones that correspond to such horizontal lines have a local maximum and a local minimum of $A(\eta,\eta_o)$. However, light cones that correspond to horizontal lines drawn below the horizontal line shown have no local extrema, since these lines do not intersect $f(\eta)$.}
\label{fig:fconcave}
\end{figure}

\pf
Consider first the case where there is a unique local maximum on each past-directed light cone and $H$ is the union of these maximal area surfaces.
Again denote the value of $\eta$ that maximizes $A(\eta,\eta_o)$ for a given $\eta_o$ by $\eta_\mrm{max}(\eta_o)$.
The only way that $\eta_\mrm{max}(\eta_o)$ could \emph{not} be a global maximum of area is if there was some $\eta_M < \eta_\mrm{max}(\eta_o)$ such that $A(\eta_M,\eta_o) > A(\eta_\mrm{max}(\eta_o),\eta_o)$.
However, for this to be possible, there must be a local minimum of $A$ in between $\eta_M$ and $\eta_\mrm{max}(\eta_o)$.
In other words, the function $f(\eta)$ must intersect the horizontal constant-$\eta_o$ line once for the local maximum, once for the local minimum, and then possibly an additional even number of times for pairs of inflection points---there cannot be more intersections if $\eta_\mrm{max}(\eta_o)$ is the unique local maximum of area.
This means that the graph of $f(\eta)$ must be concave up or concave down (\Fig{fig:fconcave}), in which case there will be some horizontal $\eta_o$ lines that do not intersect the graph of $f(\eta)$, which contradicts the requirement that $H$ have leaves on every light cone.
Therefore, $\eta = \eta_\mrm{max}(\eta_o)$ is in fact a global maximum of $A(\eta,\eta_o)$.

Then, according to Proposition~\ref{RWcont1}, the other case is where some light cones have multiple local maxima of $A(\eta,\eta_o)$, in which case $\mathcal{I}$ is only semi-infinite.
This can only happen for light cones whose tips are near the finite endpoint of the interval $\mathcal{I}$.
Beyond some threshold value of $\eta$ in the direction in which $\mathcal{I}$ is unbounded, $f(\eta)$ must still be monotonic in order for there to be leaves on every light cone.
Therefore, for $\eta_o$ beyond the threshold, the first case applies, and when there are multiple local maxima of area on a given light cone for $\eta_o$ between the threshold and the finite endpoint of $\mathcal{I}$, at least one of them is a global maximum of $A(\eta,\eta_o)$.
\eop

\bibliographystyle{utphys}
\bibliography{GSL-CNH_refs}

\end{document}